%% file: main.tex
\newif\ifconfver
\confvertrue        %declaring conference version true

\newif\ifonecoltab
\onecoltabtrue        % declaring one column version table size true
\newif\ifplainver  %declare a plain version
\plainvertrue
\newif\ifplainver  %declare a plain version
\plainvertrue
\ifplainver
    \confverfalse                         %automatically disable conf. version argument if it's plain
\fi

\ifconfver
     \documentclass[10pt,journal]{IEEEtran}
\else
    \ifplainver
        \documentclass[10pt]{article}
\usepackage{cite,palatino}
\usepackage{fullpage}
    \else
        \documentclass[12pt,draftcls,onecolumn]{IEEEtran}
    \fi
\fi

\usepackage{amsmath,amsfonts,amssymb}
\usepackage{algorithmic}
\usepackage{algorithm}
\usepackage{array}
\usepackage[caption=false,font=normalsize,labelfont=sf,textfont=sf]{subfig}
\usepackage{textcomp}
\usepackage{stfloats}
\usepackage{url}
\usepackage{verbatim,bm,stmaryrd}
\usepackage{graphicx}
\usepackage{cite}
\usepackage{bbm}
\usepackage{makecell}
\usepackage{color,amsbsy}
\usepackage{setspace}
\usepackage{dsfont,mdframed}
\usepackage{multirow}
\usepackage[normalem]{ulem}
\usepackage{pifont}% http://ctan.org/pkg/pifont
\newcommand{\xmark}{\ding{56}}%
\usepackage{tikz}
\usetikzlibrary{bayesnet}
\usetikzlibrary{arrows}

\DeclareFontFamily{U}{mathx}{}
\DeclareFontShape{U}{mathx}{m}{n}{<-> mathx10}{}
\DeclareSymbolFont{mathx}{U}{mathx}{m}{n}
\DeclareMathAccent{\widecheck}{0}{mathx}{"71}

\input{def}

\begin{document}

\newcommand{\papertitle}{\bf
Learning From Crowdsourced Noisy Labels: \\ A Signal Processing Perspective
}

\newcommand{\paperabstract}{%
One of the primary catalysts fueling advances in {\it artificial intelligence} (AI) and {\it machine learning} (ML) is the availability of {massive}, curated datasets. A commonly used technique to curate such massive datasets is crowdsourcing, where data are dispatched to multiple annotators. The annotator-produced labels are then fused to serve downstream learning and inference tasks. This annotation process often creates noisy labels due to various reasons, such as the limited expertise, or unreliability of annotators, among others. Therefore, a core objective in crowdsourcing is to develop methods that effectively mitigate the negative impact of such label noise on learning tasks. This feature article introduces advances in learning from noisy crowdsourced labels. The focus is on key crowdsourcing models and their methodological treatments, from classical statistical models to recent deep learning-based approaches, emphasizing analytical insights and algorithmic developments. In particular, this article reviews the connections between signal processing (SP) theory and methods, such as identifiability of tensor and nonnegative matrix factorization, and novel, principled solutions of longstanding challenges in crowdsourcing---showing how SP perspectives drive the advancements of this field.  
Furthermore, this article touches upon emerging topics that are critical for developing cutting-edge AI/ML systems, such as crowdsourcing in reinforcement learning with human feedback (RLHF) and direct preference optimization (DPO) that are key techniques for fine-tuning large language models (LLMs).

}

%--------

\ifplainver

    \date{\today}

    \title{\papertitle}

    \author{Shahana Ibrahim$^\star$,~Panagiotis A. Traganitis\thanks{S. Ibrahim and P. A. Traganitis contributed equally.
    S. Ibrahim is with the Department of Electrical and Computer Engineering, University of Central Florida, Orlando, FL 32816, United States. Email: shahana.ibrahim@ucf.edu. P. A. Traganitis is with the Department of Electrical and Computer Engineering at Michigan State University, East Lansing, MI 48824, United States. Email: traganit@msu.edu. X. Fu is with the School of Electrical Engineering and Computer Science, Oregon State University, Corvallis, OR 97331, United States. E-mail: xiao.fu@oregonstate.edu. G. B. Giannakis is with the Department of Electrical and Computer Engineering, University of Minnesota, Minneapolis, MN 55455, United States. Email: georgios@umn.edu. $^\star$Equal contribution.
    },~~Xiao Fu, and Georgios B. Giannakis   
    }

	\date{}

    \maketitle
    
    \begin{abstract}
    \paperabstract
    \end{abstract}

\else
    \title{\papertitle}

    \ifconfver \else {\linespread{1.1} \rm \fi

\author{Shahana Ibrahim$^\star$, Panagiotis A. Traganitis$^\star$, Xiao Fu, Georgios B. Giannakis
	
	\thanks{

	}
}

\maketitle

\ifconfver \else
\begin{center} \vspace*{-2\baselineskip}
	%11th Revision, \today \\[2\baselineskip]
\end{center}
\fi

    \ifconfver \else \IEEEpeerreviewmaketitle} \fi

 \fi

\ifconfver \else
    \ifplainver \else
        \newpage
\fi \fi

\section{Introduction}
Artificial intelligence (AI) and machine learning (ML) have made significant strides over the past few decades, expanding the potential of {learning models and algorithms}. These advancements have revolutionized natural language processing (NLP) by enhancing language generation and understanding, transformed computer vision with superior image recognition and analysis, and enabled complex decision-making capabilities.
A key driving factor behind such successes of AI/ML is the availability of \emph{large-scale, accurately labeled} training data. 
In fact, data annotation has become an indispensable integral part of the AI industry.
In a recent market report from {\it Grand View Research}, it was stated that ``the global data collection and labeling market size was valued at \$2.22 billion in 2022, and it is expected to expand at a compound annual growth rate of 28.9\% from 2023 to 2030" \cite{grandview2022}.

A prominent data annotation paradigm is crowdsourcing.
In crowdsourced data annotation systems, data items are dispatched to a group of annotators for labeling.
For each item, multiple annotators provide their individual, possibly noisy, annotations.
These annotations are then integrated to improve the overall label accuracy. 
Leveraging {such diverse annotations makes sense in data labeling,} 
as the label accuracy provided by individual annotators is sensitive to many factors. {If  annotators are human workers, their annotation accuracy is limited by their expertise level, background knowledge, and personal experience;
when annotations are provided by automated machine annotators, accuracy is affected by their model expressiveness and the quality and amount of their training data.}
The idea of systematic crowdsourcing can be dated back to more than a century ago; see the inserted box ``{Crowdsourcing in history}.''
Over the past two decades, many commercial crowdsourcing platforms have been used widely, including notable ones such as Amazon Mechanical Turk (AMT) and CrowdFlower \cite{donna2013beyond}. These platforms were instrumental in creating many highly influential datasets in modern AI history. For example, the ImageNet dataset, 
that has driven significant advances in computer vision and AI, contains approximately 14 million annotations provided by approximately 25,000 AMT annotators. More recently, large language models (LLMs), such as ChatGPT, Gemini, and Llama, used substantial amounts of crowdsourced labels {in LLM alignment
{tasks}, including supervised fine-tuning and {preference-based optimization (e.g., reinforcement learning from human feedback (RLHF) and direct preference learning (DPO))}} \cite{zhao2023survey}.

Beyond AI, crowdsourcing has also left significant footprints in a gamut of real-world applications in science and engineering.
Popular citizen science projects such as GalaxyZoo{\cite{bamford2009galaxy}} and eBird {\cite{sullivan2009ebird}} are byproducts of crowdsourcing, 
which provided effective and economical solutions to the ever-lingering data scarcity problem in life and physical sciences. Crowdsourcing techniques have also shown great promise in medical imaging-based diagnosis, producing reliable results with minimal expert supervision~\cite{biomedlabelfusion}.
In addition, {crowdsourcing is widely utilized in sensing applications through \emph{crowdsensing}~\cite{crowdsensing}, where a large group of mobile users or devices collectively share real-time data.  An example of crowdsensing is the crowd-contributed real-time information on road conditions, which is now widely available in} applications such as Google Maps and Waze. Various data fusion applications in remote sensing and healthcare that aggregate information from multiple sources of varying reliability also have strong connections to crowdsourcing techniques \cite{lahat2015multimodal}.

Despite the remarkable achievements of crowdsourcing, the fundamental challenge is that annotator-provided labels can be substantially noisy, and such noisy labels are detrimental to downstream tasks' performance. In deep learning systems for instance, noisy labels cause overfitting to noise and poor generalization. 
Early approaches to alleviating the negative impacts of crowdsourced label noise focused on improving the quality of annotators' responses during label collection. These approaches include various project management-based strategies, such as designing proper labeling workflow, querying proper experts, enhancing supervision mechanisms, and using effective incentive methods---see \cite{shah2016double} and  references therein. However, implementing such complex quality control mechanisms has become increasingly difficult and less cost-effective as data volume increases. 
For massive data annotation tasks faced by modern crowdsourcing systems,
automated annotation integration and label correction algorithms advocated by the ML community are much more relevant. In this article, the term ``crowdsourcing'' refers to this type of automated systems and algorithms, unless otherwise specified.

\noindent
{\bf Our Goal.} 
In this feature article, we aim to provide insights into key developments of learning from crowdsourced noisy labels {from a signal processing (SP) vantage point}. {Particularly, we will offer an insightful review of the impactful works in this domain. These include key crowdsourcing models, ranging from classical statistical models to recent deep learning-based crowdsourcing models. Major emphasis will be placed on analytical and algorithmic developments behind the success of these models.} As one will see, designing crowdsourcing algorithms and frameworks presents a series of challenges in problem distillation, optimization, and performance characterization. Interestingly---but not very surprisingly---many design considerations and solutions in crowdsourcing are deeply intertwined with theory and methods of SP.
{Therefore, we will emphasize how the unique strengths of the signal processing society (SPS) in optimization, statistics, and advanced (multi-)linear algebra, among others, have propelled crowdsourcing research and still remain significantly relevant to address modern-day challenges.} {We  focus on presenting representative works,
highlighting key ideas in modeling, learning criterion formulation, algorithm
design, and theoretical advancements, rather than presenting an exhaustive survey of crowdsourcing
algorithms.
To the best of our knowledge, such a tutorial treatment that aims to balance between review of developments and technical depth has been elusive.}

\begin{mdframed}
[backgroundcolor=gray!10,topline=false,
	rightline=false,
	leftline=false,
	bottomline=false]
{\bf Crowdsourcing in history.}
The term ``crowdsourcing" was first introduced by business journalist Jeff Howe in his 2006 \textit{Wired} magazine article on task outsourcing to Internet users, yet the concept has deep historical roots.
One of the earliest examples of a successful crowdsourcing project is the compilation of the Oxford English dictionary back in the nineteenth century, where hundreds of English speaking readers' efforts were utilized to collect words and their meanings. Years later in 1907, a statistical perspective to this paradigm---``the wisdom of the crowd"---was first postulated by the famous statistician Sir Francis Galton, when he observed a crowd at an auction accurately predicting the weight of an ox through their collective guessing.  
In the late 1990s, many large-scale annotation projects, such as TreeBank, FrameNet, and PropBank,
began to emerge, driven by the need for large amounts of curated data in NLP tasks. Another notable early effort in crowdsourcing was pioneered by Luis von Ahn in 2004, who introduced an online game to generate image annotations, which sparked substantial interest in utilizing online workers for large-scale annotation tasks. 
\end{mdframed}

\noindent
{\bf Notation.} We use $x$, $\bm{x}$, and $\bm{X}$ to denote scalar, vector, and matrix, respectively; 
    both $[\bm{x}]_i$ and $\bm x(i)$ refer to the $i$th entry of vector $\bm{x}$;
    $[\bm X]_{i,j}$ or $\bm X(i,j)$ is the entry in the $i$th row and $j$th column of $\bm{X}$; $\bm X(i,:)$ or $[\bm X]_{i,:}$ denote the $i$th row of $\bm X$; $\bm X(:,j)$ or $[\bm X]_{:,j}$ the $j$th column of $\bm X$; $\mathbb{I}[A]$ is the indicator function, i.e., $\mathbb{I}[A]=1$, if the event $A$ occurs, otherwise $\mathbb{I}[A]=0$; 
$\bm{I}_K$ denotes the identity matrix of size $K$; 
$\|\bm x\|_2$ denotes the $\ell_2$ norm; $\|\bm X\|_{\rm F}$ and $ \|\bm X\|_2$ denote the Frobenius and the spectral norm of $\bm{X}$, respectively; 
$[N]:=\{1,\ldots,N\}$ is the set of natural numbers from 1 to $N$; ${\rm krank}(\bm X)$ denotes the Kruskal rank of the matrix $\bm X$; ${\rm trace}(\bm X)$ is the trace, i.e., the sum of the diagonal entries, of matrix $\bm X$; 
$\circ$ denotes the outer-product; $\prob$ denotes probability or probability mass function (PMF) or probability density function (PDF), and $\prob(X;\bm{\theta})$ denotes a distribution parametrized by $\bm{\theta}$; {$\text{w.p.}$ stands for ``with probability."}

\section{Problem Settings} 
\label{sec:label_integration} 
In this section, we first state the {label integration problem in} crowdsourcing. Next, we introduce two crowdsourcing-based system designs that leverage label integration.

\subsection{Problem statement} \label{ssec:problem_statement}
As mentioned, crowdsourcing can be used to integrate various types of annotations (e.g., class labels \cite{dawid1979maximum}, traffic information \cite{tong2017spatial}, and bird counts); that is, in principle, the annotations in crowdsourcing can be {either discrete or continuous values}. We will henceforth use categorical class label annotations as the primary working example to introduce crowdsourcing ideas and algorithms.
Consider a dataset consisting of $N$ data items $\mathcal{X}:=\{\bm{x}_n\}_{n=1}^{N}$, where $\bm{x}_n \in \mathbb{R}^{D}$ is the feature vector representing the $n$-th data item that belongs to one of $K$ classes. Let $\{y_n\}_{n=1}^{N}$ denote the set of \textit{ground-truth} labels, where $y_n \in [K],~\forall n$; i.e., $y_n = k$ if $\bm{x}_n$ belongs to class $k$. 
The ground-truth labels $\{y_n\}_{n=1}^{N}$ are \textit{unknown}. Assume that $M$ crowdsourced workers (annotators) are employed to label the dataset $\mathcal{X}$, i.e., to give their estimates of $\{y_n\}$. 
We use $\check{y}_n^{(m)}\in[K]$ to denote the label assigned to the $n$th data item by the $m$th annotator.
Given annotations 
$\{\check{y}_n^{(m)}, m=1,\ldots,M\}_{n=1}^{N}$,  \emph{label integration} aims to find estimates $\{\hat{y}_n\}_{n=1}^{N}$ and reliabilities $\{ {\rm Pr}(\check{y}_n^{(m)}|y_n), m=1,\ldots,M\}_{n=1}^N$, that are subsequently employed along with features
$\{\bm{x}_n\}_{n=1}^{N}$
for downstream learning and inference, as depicted in Fig. 1.   
{In practice, each annotator typically labels only a subset of ${\cal X}$ as requiring every annotator to label all data items would be prohibitively expensive.}
Annotators may be humans, pre-trained ML algorithms, or predefined decision rules. Crowdsourced annotations are thus typically noisy, which means that $\widecheck{y}_n^{(m)} \neq y_n$ for many $(m,n)$ pairs.

    {Note that the problem setup is presented in its ``basic" form for clarity and ease of exposition. Nevertheless, complex scenarios may arise. For instance, the notion of a well-defined, objective ground-truth does not always hold, especially in domains such as medical diagnosis, sentiment analysis, and artistic evaluation---where labels could be fairly subjective. In addition, various data/label availability situations may occur; e.g., occasionally, a limited number of ground-truth labels may be available, leading to semi-supervised problem settings. 
}

\begin{figure}[t]
    \centering
    \includegraphics[scale=0.45]{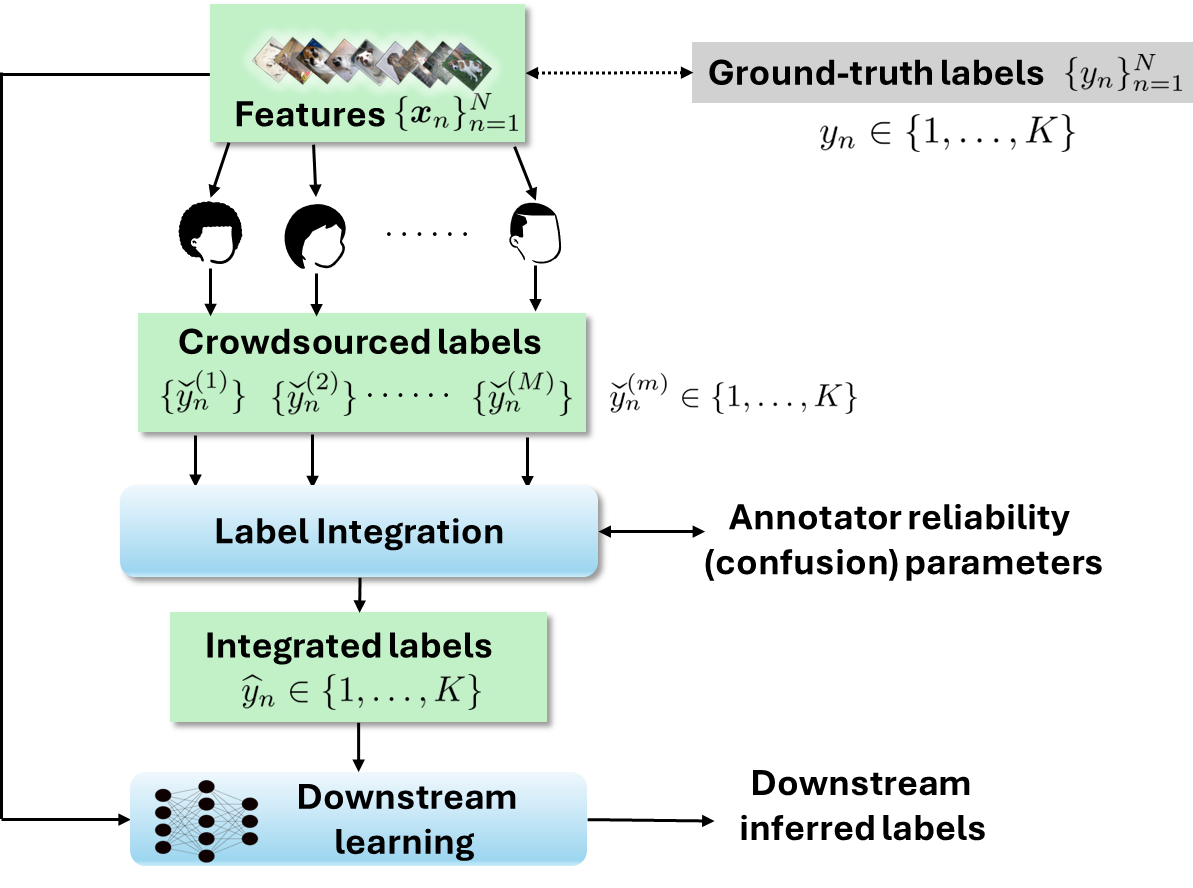}
    \caption{The two-stage strategy with label integration and annotator confusion estimation followed by {downstream learning}.} 
    \label{fig:single_stage}
\end{figure}

\begin{figure}[t]
    \centering
    \includegraphics[scale=0.45]{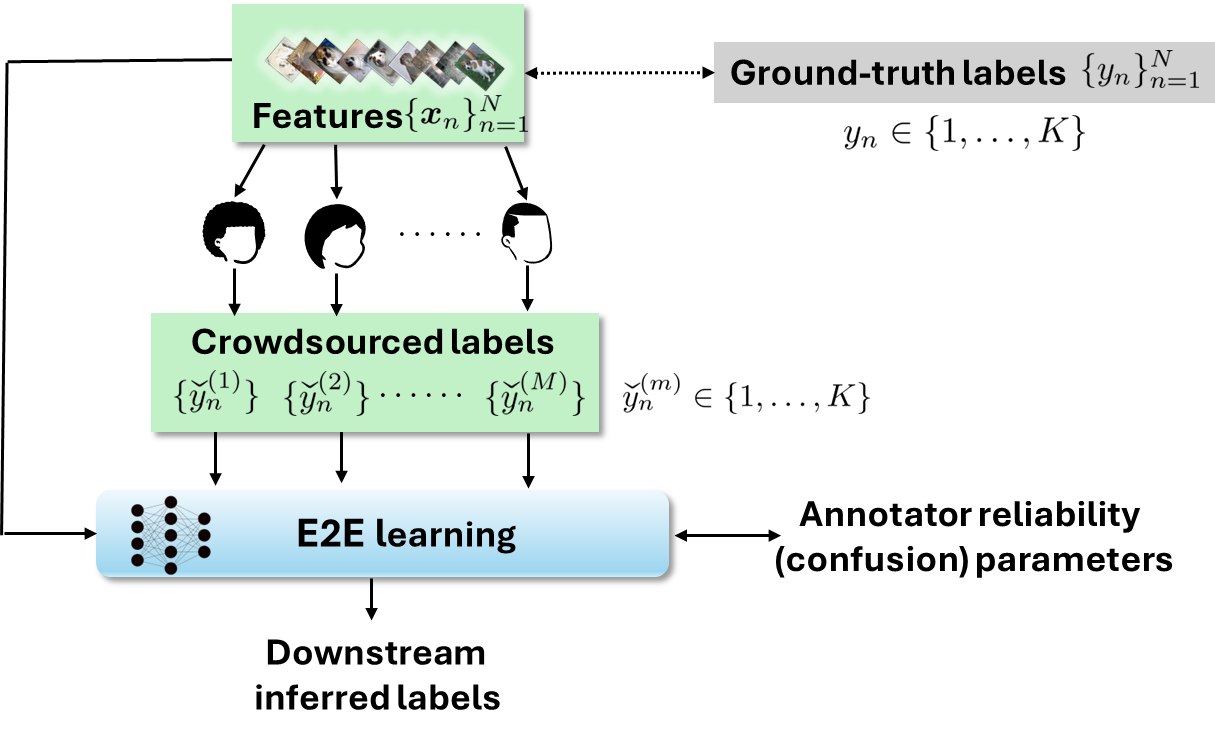}
    \caption{The E2E learning strategy for joint estimation of the ground-truth label predictor and annotator confusions.} 
    \label{fig:e2e}
\end{figure}

\subsection{Crowdsourcing systems: Two paradigms}
\noindent
Crowdsourcing systems can be categorized either as label integration approaches (see, e.g., \cite{dawid1979maximum,traganitis2018blind,zhang2014spectral,ibrahim2019crowdsourcing}), or, as end-to-end (E2E) learning approaches (e.g., \cite{ibrahim2023deep,rodrigues2018deep,tanno2019learning,raykar2010learning}). The former category isolates the label integration module from the downstream learning and inference module, while E2E directly uses annotations 
$\{\check{y}_n^{(m)}, m=1,\ldots,M\}_{n=1}^{N}$
to  train the downstream module.

\noindent
{\bf Label integration.}
Fig.~\ref{fig:single_stage} depicts the first paradigm of a crowdsourcing system, where designing the label integration module  does not account for the ensuing module of the downstream task {(e.g., training an image classifier using the integrated labels)} at hand. Annotations $\{\check{y}_n^{(m)}, m=1,\ldots,M\}_{n=1}^{N}$
are fused to obtain the estimates
$\{\widehat{y}_n\}_{n=1}^N$  in the hope that
$\{\widehat{y}_n = y_n\}_{n=1}^N$. The
reliability of annotators, referred to as ``annotator confusions'' in Fig. ~\ref{fig:single_stage}, can be also estimated in the process.
Estimates $\{\widehat{y}_n\}_{n=1}^N$ are then used as `known' labels to train the downstream module. 

\noindent
{\bf End-to-End (E2E) learning.}
Fig.~\ref{fig:e2e} illustrates a typical E2E crowdsourcing system, in
which both features and noisy labels are used as input to directly train learning systems for target downstream tasks. As there is no stage breakdown, these systems are called end-to-end (E2E) systems. When the target is to train a classifier $\widehat{\bm f}_{\bm \theta}: \mathbb{R}^D \rightarrow \mathbb{R}^K$, {parameterized by ${\bm{\theta}}$,}  E2E systems design a loss function $\ell$ such that
\[      \widehat{\bm f}_{\bm \theta} \leftarrow \arg\min_{\bm \theta,\bm \eta}~{\ell}(\{ \x_n\},\{\widecheck{y}_n^{(m)}\},\bm \theta,\bm \eta)  \]
where $\bm \eta$ represents additional model parameters, {such as annotator reliability (confusion) parameters,} according to specific loss designs. 
{The learned $\widehat{\bm f}_{\bm \theta}$ is expected to be a reliable {ground-truth} label predictor.}
For example, let the function $\bm f^\star:\mathbb{R}^D\rightarrow \mathbb{R}^K$ 
denote the ground-truth label posterior distribution, whose $k$-th entry is 
\begin{align}\label{eq:gt_pedictor}
    [\bm f^\star(\x_n)]_k ={\sf Pr}(y_n=k|\x_n),\quad \forall (\x_n,y_n).
\end{align}
The goal of several E2E approaches (see, e.g., \cite{rodrigues2018deep,ibrahim2023deep,tanno2019learning}) is to learn $\widehat{\bm f}_{\bm \theta}$ so that $\widehat{\bm f}_{\bm \theta} \approx \bm f^\star$.

\medskip
{
Generally speaking, label integration approaches can afford lightweight and tractable algorithms, as they do not require training a learning system $\bm f_{\bm \theta}$ that is usually parameterized by a complex function class, such as kernels or neural networks. On the other hand, E2E methods often exhibit more appealing performance, possibly because  they naturally leverage information from data features \emph{jointly} with the underlying structure of the specific tasks.}  {Both crowdsourcing paradigms can be considered \emph{unsupervised}, as {they can work without} any ground-truth labels available to guide the learning process.}
The developments in label integration and E2E learning based crowdsourcing are reviewed in Sec.~\ref{sec:label_integration} and Sec.~\ref{sec:e2e}, respectively.

\section{Label Integration Approaches}   \label{sec:label_integration}
This section deals with designing the label integration module in Fig.~\ref{fig:single_stage}. {Widely used models for label integration are introduced, followed by algorithms to estimate their parameters. Next, models and algorithms tailored to specific scenarios are outlined, concluding with theoretical considerations.}

\subsection{Majority voting and challenges}
{The arguably simplest} approach to label integration is \emph{majority voting} (MV), where the estimated label $\widehat{y}_n$ is the one voted by most annotators; that is, with $\mathbb{I}$ denoting the indicator function, it holds that  
\begin{equation}
    \label{eq:MV}
        \widehat{y}_n = \arg\max_{k}\sum_{m=1}^{M}\mathbb{I}[\widecheck{y}_n^{(m)} = k].
\end{equation}
However, MV is not always effective, as it implicitly assumes equal reliability {and statistical independence} across annotators. 
Furthermore, due to the significant time and effort required, it is not economical to query all annotators to label every 
data item. This means that many data may not have sufficient annotations that merit a voting process. Under such circumstances, MV can be far from optimal for label integration \cite{karger2014budget}. 
To account for unequally reliable {annotators}, \emph{weighted majority voting} (WMV) schemes were proposed; see e.g.,~\cite{littlestone1994weighted}, where a non-negative scalar weight $w^{(m)}$ capturing the labeling accuracy of annotator $m$ is introduced, {leading} to  
    \begin{equation}
    \label{eq:weighted_MV}
        \widehat{y}_n = \arg\max_{k}\sum_{m=1}^{M}w^{(m)}\mathbb{I}[\widecheck{y}_n^{(m)} = k].
    \end{equation}
{Note that accurate weight estimation {\it per se} is a challenging task.} {Nevertheless, the simplicity of MV schemes make them popular baselines for label integration tasks.}

\subsection{Label noise models}
    \label{ssec:key_crowd_models}

A milestone in {label integration} research occurred when MV approaches were superseded by probabilistic label noise models. Relying on such generative models, these approaches can assess the performance of {annotators}. We will outline 
a number of representative models in this genre {as follows}:
    \begin{figure}[t]
    \centering
    \includegraphics[scale=0.45]{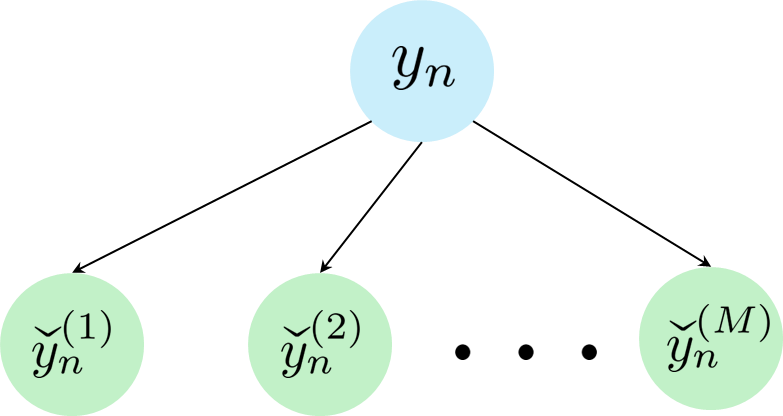}
    \caption{DS graph model \cite{dawid1979maximum}. Green colored circles indicate observed variables, whereas the blue one is the latent variable. 
    }
    \label{fig:DS}
\end{figure}

\noindent
{\bf Dawid-Skene (DS) model.}
The seminal work of Dawid and Skene introduced one of the most influential probabilistic models for crowdsourcing  in the late 1970s \cite{dawid1979maximum}. 
{Under the DS model, the ground-truth label $y_n$ is a latent discrete random variable, taking values from $1,\dots, K$.} 
{The responses across annotators $\widecheck{y}_n^{(1)}, \dots,\widecheck{y}_n^{(M)}$ are observed random variables.}
The key assumption here is that, given the ground-truth label, annotators' responses are conditionally independent, {i.e.,
${\sf Pr}(\widecheck{y}_n^{(1)}=k_1,\dots, \widecheck{y}_n^{(M)}=k_M | y_n = k) =  \prod_{m=1}^M {\sf Pr}(\widecheck{y}_n^{(m)}=k_m|y_n=k)$.}
In other words, the DS model is a \textit{naive Bayes} model; see Fig.~\ref{fig:DS}. 

The joint probability of all annotator responses {for the $n$-th data item} can be expressed as
    \begin{align}
        {\sf Pr}(\widecheck{y}_n^{(1)}=k_1,\dots, \widecheck{y}_n^{(M)}=k_M) = \sum_{k=1}^K \prod_{m=1}^M {\sf Pr}(\widecheck{y}_n^{(m)}=k_m|y_n=k){\sf Pr}(y_n=k)  \label{eq:DSmodel}
\end{align}
where $k_1,\dots,k_M \in [K]$ are the annotator responses, and we have used the law of total probability, Bayes rule, and the conditional independence of annotator responses given the ground-truth label. {Data under the DS model are assumed to be drawn i.i.d. from some unknown distribution, therefore the probabilities ${\sf Pr}(y_n=k)$ and  ${\sf Pr}(\widecheck{y}_n^{(m)}=k_m|y_n=k)$ do not depend on the data index $n.$} Let  
\begin{align}\label{eq:confusionmatrix}
        \bm A_m(k',k) \triangleq {\sf Pr}(\widecheck{y}_n^{(m)}=k'|y_n=k),\;\;\; \bm d(k) \triangleq {\sf Pr}(y_n=k), ~\forall k,k' \in [K],
\end{align}
where $\bm A_m \in \mathbb{R}^{K \times K}$ is termed the {\it confusion matrix of} annotator $m$ and $\bm d \in \mathbb{R}^K$ {is the PMF of the ground-truth label distribution.} As the name suggests, {the off-diagonal elements of $\bm A_m$ characterize the probabilities of annotator $m$ making mistakes}; an ideal annotator would have $\bm A_m = \bm I_K.$ Also note that, as each column of a confusion matrix is a PMF, it holds that $\bm A_m\geq \bm 0, \mathbf{1}^{\top}\bm A_m = \mathbf{1}^{\top},$ with $\mathbf{1}$ denoting the all-ones vector of appropriate dimension. Thus, both $\bm d$ and columns of $\bm A_m$ belong to the so-called  $(K-1)$-dimensional probability simplex, {$
\bm \varDelta_K \triangleq \{\bm p \in \mathbb{R}^{K}: \bm p \geq \bm 0, \mathbf{1}^{\top}\bm p = 1\}.$} {Accordingly, annotator responses $\widecheck{y}_n^{(m)}$ are random variables drawn independently from a categorical distribution parametrized by the vector $\bm A_m \bm d$, that is, $\widecheck{y}^{(m)}_n \sim {\rm categorical}(\bm A_m \bm d)$. Furthermore, given $y_n=k$, annotator responses are drawn from a categorical distribution parametrized by the $k$-th column of the corresponding confusion matrix, i.e., $\widecheck{y}_n^{(m)} | y_n = k \sim {\rm categorical}(\bm A_m(:,k))$. }

{Under the DS model, if parameters $\bm A_m$'s and $\bm d$ are known, one can construct optimal ground-truth label estimators using the \textit{maximum a posteriori} (MAP) principle, {that is, by} maximizing the posterior distribution of the unknown label, given annotator responses:}
\begin{align}
                \widehat{y}_n & = \arg\max_{k\in [K]}{\prob(y_n = k | \widecheck{y}_n^{(1)} = k_1,\ldots, \widecheck{y}_n^{(M)} = k_M) }\notag \\ &=\arg\max_{k\in [K]}{\prob(\widecheck{y}_n^{(1)} = k_1,\ldots,\widecheck{y}_n^{(M)} = k_M | y_n = k)\prob(y_n = k)}
                \notag \\ &  = 
          \arg\max_{k \in [K]} \log\bm d(k) 
          + \sum_{k'=1}^{K}\sum_{m=1}^{M}\mathbb{I}[\widecheck{y}_n^{(m)} = k']\log\bm A_m(k',k),
                  \label{eq:label_integration_DS}
\end{align}
where we have successively used Bayes' rule and the definitions in \eqref{eq:confusionmatrix}. 
{Nevertheless, DS model parameters are not known \textit{a priori}. }  Hence,  label integration  under the DS model amounts to estimating $\{ \bm A_m\}$ and $\bm d$ accurately {from the available crowdsourced labels $\{\widecheck{y}_n^{(m)}\}$. The total number of parameters of the general DS model is $O(MK^2)$, suggesting that at least $\Omega(MK^2)$ annotations are required to estimate $\{ \bm A_m\}_{m=1}^M$ and $\bm d$.}

\begin{mdframed}
[backgroundcolor=gray!10,topline=false,
	rightline=false,
	leftline=false,
	bottomline=false]
{\bf Links with communications, information theory, and SP.}
The DS model can be viewed as a \textit{single-input multiple-output} (SIMO) channel with input the per datum ground-truth label $y_n$, and outputs the $M$ annotator responses $\{\widecheck{y}_n^{(m)}\}_{m=1}^{M}$, while the 
matrices  $\{ \bm A_m\}_{m=1}^M$ can be associated with the unknown ``channel'' characteristics. A setup similar to crowdsourcing is that of \textit{decentralized detection}~\cite{tsitsiklis_dd}, where a distributed network of sensors (annotators) observe the same environmental phenomenon, and send their  observations (labels) to a fusion center, and the fusion center has to recover the underlying hypothesis of the environment. {The key difference between decentralized detection and crowdsourcing is that in the former case, the fusion center can control the characteristics of both the 
annotators and the fusion rule.} 
{The crowdsourcing setting with real-valued annotations also bears resemblance to SP  problem of \emph{blind multichannel deconvolution}~\cite{multichannel_blind_deconv}, where the unobserved input and the channel are inferred from multiple noisy observed signals. 
Further, crowdsourcing bears similarities to the \textit{chief executive officer} (CEO) problem in information theory \cite{berger1996ceo}, where $M$ agents observe noisy sequences and the CEO aims to recover the ground truth under a communication constraint, similar to the labeling cost budget in crowdsourcing.  {Crowdsourcing was also shown to have interesting connections with other SP techniques, e.g., coding theory  \cite{vempaty2014reliable} and \textit{graph signal processing} (GSP) \cite{maroto2018efficientworkerassignmenGSP}.}}
\end{mdframed}

\noindent
{\bf Special DS models.}
 The DS model serves as a foundation for different {and simpler} parametric models that are frequently employed. The so-called \emph{one-coin} model \cite{ghosh2011moderates} is a simplified DS model that encodes each annotator's reliability using only one parameter, that is 
 \begin{align*}
    \prob(\widecheck{y}_n^{(m)} = k' | y_n = k) = \begin{cases}
        p_m, & k'=k\\
        \frac{1-p_m}{K-1}, & k' \neq k
    \end{cases}, 
 \end{align*}
where annotator $m$ determines responses with a single biased coin flip. Per data item $n$, annotator $m$ assigns the correct label with probability $p_m$, and
a wrong label with equal probabilities across the remaining $K-1$ classes. 
Other popular DS model variants include the \textit{spammer-hammer} model \cite{karger2011budget} and the \emph{confusion vector} model \cite{minimax_entropy}. In the spammer-hammer model, each annotator is a ``hammer'' with probability $q$, providing correct labels, or a ``spammer'' with probability $1-q$, providing random labels. {Both {the} one-coin and spammer-hammer models {work under fairly restrictive conditions}, as the annotator confusion probabilities are {assumed to be} the same irrespective of their ground-truth class {labels}.}
The confusion vector model {{generalizes the setting} by employing} a parameter vector $\bm a_m \in \mathbb{R}^K$, per annotator $m$, to characterize reliabilities as 
\begin{align*}
    \prob(\widecheck{y}_n^{(m)} = k' | y_n = k) = \begin{cases}
        \bm a_m(k), & k^{'}=k\\
        \frac{1-\bm a_m(k)}{K-1}, & k' \neq k
    \end{cases}.
 \end{align*}

These special DS models offer succinct characterizations of annotator confusions{; the one-coin and spammer-hammer models involve $M$ parameters, while the confusion vector model includes $O(MK)$ parameters.}
{They are less general than the DS model, but can afford reduced-complexity parameter estimators with quantifiable performance guarantees; see, e.g., \cite{ghosh2011moderates,dalvi2013aggregating,karger2011budget}. }{The reduced number of parameters is crucial in situations where the number of available annotations is relatively small.}

\noindent
{\bf Incorporating item difficulties.}
The DS model introduced in \cite{dawid1979maximum} assumes that  annotator behavior is invariant across all data, that is, $\bm A_m$ is the same for all $n$. 
{Nonetheless, it may be more realistic to assume that the difficulty of labeling varies across data items.}
To capture both annotator behavior and item difficulty, 
{the confusion probability can be modeled as follows
\cite{zhou2014aggregating}:}
\begin{equation*}
        \prob(\widecheck{y}_n^{(m)} = k'|y_n=k ) = \frac{\exp(\bm A_m(k',k)+\bm B_n(k',k))}{\sum_{k'} \exp(\bm A_m(k',k)+\bm B_n(k',k))}
\end{equation*}
where $\bm A_m \in \mathbb{R}^{K \times K}$ is defined as before, and $\bm B_n \in \mathbb{R}^{K \times K}$ is an item-specific confusion matrix that reflects  item difficulty. Along the same spirit, the one-coin model can be also generalized to incorporate item difficulty; see the \textit{generative model of labels, abilities, and difficulties} (GLAD) model in \cite{GLAD}. {As item difficulty is taken into account, these models increase the number of parameters to be estimated.}

\noindent
{\bf Bayesian models.}  
{To incorporate prior information, reduce the number of parameters, and enhance model interpretability, Bayesian models were also introduced on top of the DS model (see, e.g., \cite{raykar2010learning,BCC_Kim}). 
Using the fact that both $\bm d$ and $\bm A_m(:,k)$ are PMFs}, \cite{BCC_Kim} imposes Dirichlet priors on $\bm d$ and $\bm A_m(:,k)$'s, while \cite{raykar2010learning} assumes a Beta prior for the confusion parameters in the binary classification case.

\subsection{{Learning the label noise models}} \label{ssec:methods_label_int}  
{Under the aforementioned noise generation models, label integration boils down to learning the key model parameters}, e.g., the confusion matrices and the prior probability vector in the DS model. In this subsection, we briefly review some representative methods for learning these models.

\noindent
{\bf Expectation Maximization (EM).} The seminal work by Dawid and Skene \cite{dawid1979maximum} sought a maximum-likelihood estimator (MLE) of annotator confusion matrices and class priors of the DS model. Collect all ground-truth labels and annotator responses in ${\cal Y} =\{y_n\}$ and $\widecheck{\cal Y} = \{\widecheck{y}_n^{(m)}\}$, respectively, and all unknown model parameters in $\bm \psi = \{\bm A_1,\dots, \bm A_M, \bm d\}$. Then, the MLE of $\bm \psi$ is given by 
{ 
\begin{align} \label{eq:mle_psi}
        \widehat{\bm \psi}  &= \underset{\bm \psi}{\text{arg~max}}~\log {\sf Pr}(\widecheck{\cal Y};\bm \psi) =\underset{\bm \psi}{\text{arg~max}}~ \sum_{n=1}^N \log \prob(\widecheck{y}_n^{(1)},\dots,\widecheck{y}_n^{(M)};\bm \psi) \nonumber\\
        &= \underset{\bm \psi}{\text{arg~max}}~ \sum_{n=1}^N \log \sum_{k=1}^K \bm d(k) \prod_{m=1}^M \sum_{k'=1}^K\mathbb{I}[\widecheck{y}_n^{(m)}=k']\bm A_m(k',k),
\end{align} 
}where $\log {\sf Pr}(\widecheck{\cal Y};\bm \psi)$ is the log-likelihood function of observed annotator labels, parametrized by $\bm \psi$. As directly optimizing the log-likelihood is often intractable, Dawid and Skene introduced an EM-based algorithm to learn the DS model parameters. {The EM algorithm is the workhorse for learning naive Bayes and mixture models.} 
 Using the EM algorithm, the log-likelihood function is maximized  iteratively, with the following steps performed at each iteration:

{\it (i) The ``Expectation" (E) step}: 
The E-step in iteration $t$ is performed as follows:  
        \begin{align*}
        &Q(\bm \psi; \bm \psi^t) = \mathbb{E}_{{\cal Y}\sim {\sf Pr}({\cal Y};\widecheck{\cal Y}, \bm \psi^t)}[\log {\sf Pr}(\widecheck{\cal Y},{\cal Y}; \bm \psi)] = \sum_{n=1}^N \sum_{k=1}^K q(y_n=k; \bm \psi^t)  \log {\sf Pr}(y_n = k,\widecheck{y}_n^{(1)},\dots,\widecheck{y}_n^{(M)};\bm \psi),
\end{align*}
where the superscript $t$ denotes iteration index, and the expectation is taken with respect to (w.r.t.) the posterior probability ${\sf Pr}({\cal Y};\widecheck{\cal Y}, \bm \psi^t)$ based on the current estimates $\bm \psi^t=\{\bm A^t_1,\dots, \bm A^t_M, \bm d^t\}$. It turns out that $Q(\bm \psi; \bm \psi^t)$ is essentially a lower bound of the log-likelihood. The E-step boils down to  estimating $q(y_n=k; \bm \psi^t)={\sf Pr}(y_n=k| \widecheck{y}_n^{(1)},\dots,\widecheck{y}_n^{(M)};\bm \psi^t)$ given $\bm A^t_m$'s and $\bm d^t$. {Under the DS model and using Bayes rule, $q(y_n=k; \bm \psi^t)$ admits a simple closed form, i.e.,}{
\begin{equation}
    \label{eq:DS_EM_posterior}
    q(y_n = k;\bm{\psi}^t) = \frac{\exp\left(\log\bm{d}^{t}(k) +\sum_{m=1}^{M}\sum_{k'=1}^{K} \log\bm{A}_m^{t}(k',k){\mathbb{I}[\widecheck{y}_n^m=k']}\right)}{\sum_{k'=1}^{K}\exp\left(\log\bm{d}^{t}(k') +\sum_{m=1}^{M}\sum_{k''=1}^{K} \log\bm{A}_m^{t}(k'',k'){\mathbb{I}[\widecheck{y}_n^m=k'']}\right)}.
\end{equation}
Note that this posterior has a similar form to the MAP estimator of \eqref{eq:label_integration_DS}.
}

{\it (ii) The ``Maximization" (M) step:} The  M-step refines model parameters by maximizing $Q(\bm \psi; \bm \psi^t)$, which also admits analytical updates:
\begin{align*} 
&\bm A^{t+1}_m (k',k)= \frac{\sum_{n=1}^N q(y_n=k; \bm \psi^t) \mathbb{I}[\widecheck{y}_n^m=k']}{\sum_{k''=1}^K\sum_{n=1}^N q(y_n=k; \bm \psi^t)\mathbb{I}[\widecheck{y}_n^m=k'']},\\
&\bm d^{t+1}(k) = \frac{\sum_{n=1}^N q(y_n=k; \bm \psi^t)}{\sum_{k'=1}^K\sum_{n=1}^N q(y_n=k'; \bm \psi^t)}.
\end{align*}
The EM scheme can also be readily adapted to learn the special cases of the DS model, i.e., one-coin, confusion vector, spammer-hammer, and GLAD models as discussed in \ref{ssec:key_crowd_models}. A similar EM strategy is also employed to learn the Bayesian model proposed in \cite{raykar2010learning}.

{One salient feature of the EM algorithm is its scalability---it enjoys linear computational complexity in $N$ and $M$, which is appealing for large-scale crowdsourcing problems. However, the EM iteration guarantees convergence only to a local optimum of the generally nonconvex MLE objective; thus, convergence to the global optimum calls for careful parameter initialization\cite{zhang2014spectral,ibrahim2021recovering}.
}

\noindent
{\bf Spectral methods.}
One of the notable spectral methods for label integration is the eigendecomposition-based approach proposed in \cite{ghosh2011moderates}  under the one-coin binary model. {Consider the binary classification problem with unobserved ground-truth labels $y_n\in \{-1,+1\}$ and the one-coin model parameters $p_m$'s, where $p_m$ is the probability that annotator $m$ provides the correct label. {Let $z_{n}^{(m)}$ be the correctness indicator of annotator $m$ on item $n$; i.e., $z_{n}^{(m)}=1$ if annotator $m$ provides the correct label and $z_{n}^{(m)}=-1$ otherwise. Then, it can be shown that
\begin{align*}
    z_{n}^{(m)}z_{n'}^{(m)} = \begin{cases}
        1, & \text{w.p.}~p_m^2+(1-p_m)^2 ,\\
        -1, & \text{w.p.}~1-p_m^2-(1-p_m)^2.
    \end{cases}
\end{align*}
Since $\widecheck{y}_n^{(m)}= z_{n}^{(m)} y_n$ holds, we further obtain:
\begin{align}
    \mathbb{E}\left[\sum_{m=1}^M \widecheck{y}_{n}^{(m)}\widecheck{y}_{n'}^{(m)}\right] &= y_n y_{n'} \mathbb{E}\left[\sum_{m=1}^M z_{n}^{(m)}z_{n'}^{(m)}\right] 
    =\begin{cases}
        y_n y_{n'}\kappa, & n \neq n' ,\\
        M, & n=n',
    \end{cases} \label{eq:svd}
\end{align}
where $\kappa=\sum_{m=1}^M (2p_m-1)^2$. By defining the annotator response matrix $\widecheck{\bm U}$ with entries $\widecheck{\bm U}(n,m)=\widecheck{y}_{n}^{(m)}$, the relation in \eqref{eq:svd} can also be expressed as
\begin{align}\label{eq:onecoin-ident}
  \mathbb{E}[\widecheck{\bm U}\widecheck{\bm U}^{\top}]= \kappa \bm y \bm y^{\top} + (M-\kappa) \bm I_{N}, 
\end{align}
where $\bm y = [y_1,\dots, y_N]^{\top}$.
Based on the above, \cite{ghosh2011moderates} proposed an intuitively simple spectral algorithm where the ground-truth labels $\bm y$ are extracted from the top eigenvector of {the empirical estimate of} $\mathbb{E}[\widecheck{\bm U}\widecheck{\bm U}^{\top}]$. Consequently, annotator confusions $\bm p = [p_1,\dots, p_M]^{\top}$ can be inferred from the estimated $\bm y$.
}
Nonetheless, the approach requires that the annotator response matrix is fully observed, meaning that all annotators provide labels for all data items. To accommodate the incomplete labeling scenarios, \cite{dalvi2013aggregating} extended this strategy by considering an annotator-item binary matrix alongside the annotator response matrix and performing a joint singular value decomposition (SVD) operation, still under the one-coin model.
{The eigendecomposition approach was later extended to the general DS model in the case of binary classification in \cite{jaffe2015estimating}.

}

\noindent
{\bf Moment-based approaches: From tensor decomposition to nonnegative matrix factorization.}
{To identify the parameters of the general DS model,}
 \cite{traganitis2018blind} proposed a moment matching approach that considered the third-order moments of annotator responses as follows}:
\begin{align}\label{eq:3rdordertensor1}
\Expect[\widecheck{\bm{y}}_n^{(m)}\circ\widecheck{\bm{y}}_n^{(i)}\circ{\widecheck{\bm{y}}_n^{(j)}}] &=  \sum_{k=1}^K \bm d(k)  \bm A_m(:,k) \circ \A_i(:,k)\circ \A_j(:,k),~\forall m\neq i\neq j,
\end{align}
where $\circ$ denotes the outer product (i.e.,
$\tX=\bm a\circ \bm b \circ \bm c \Leftrightarrow \tX(m,i,j)=\bm a(m)\bm b(i)\bm c(j)$), and $\widecheck{\bm{y}}_n^{(m)}$ denotes the $K$-dimensional one-hot encoding of the annotator response random variable $\widecheck{y}_n^{(m)}$, i.e., if $\widecheck{y}_n^{(m)}=k$, then $\widecheck{\bm{y}}_n^{(m)} = \bm e_k$, where $\bm e_k \in \mathbb{R}^K$ is a unit vector with $[\bm e_k]_k = 1$ and zeros elsewhere. {In general, the conditional independence assumption of the DS model induces the outer product expression of the higher-order moments.}
To see why, consider the second-order moment
\begin{align}\left[\Expect[\widecheck{\bm{y}}_n^{(m)}\circ\widecheck{\bm{y}}_n^{(i)}]\right]_{k_1,k_2} &= {\sf Pr}(\widecheck{y}_n^{(m)}=k_1,\widecheck{y}_n^{(i)}=k_2) \nonumber\\
&= \sum_{k=1}^K \underbrace{{\sf Pr}(y_n=k)}_{\bm d(k)}  \underbrace{{\sf Pr}(\widecheck{y}_n^{(m)}=k_1|y_n=k)}_{\bm A_m(k_1,k)}  \underbrace{{\sf Pr}(\widecheck{y}_n^{(i)}=k_2|y_n=k)}_{\bm A_i(k_2,k)},\nonumber
\end{align}
which implies $\Expect[\widecheck{\bm{y}}_n^{(m)}\circ\widecheck{\bm{y}}_n^{(i)}]=\sum_{k=1}^K\bm d(k) \A_m(:,k)\circ\A_i(:,k)= \bm A_m\diag(\bm{d})\bm A_i^{\top}$;
a similar derivation holds for the third-order moments.

Using third-order moments, a {\it coupled tensor factorization} (CTD) criterion can be used to identify $\bm d$ and $\{\A_m\}$ \cite{traganitis2018blind}:
\begin{align}\label{eq:coupledtensor}
        &\underset{{\{\bm A_m\}_{m=1}^M}, \bm d}{\rm{minimize}} \sum_{\substack{m=1\\i > m \\ j > i}}^M\|\tT_{m,i,j}-\llbracket \bm d, \bm A_m,\bm A_i,\bm A_j\rrbracket\|_{\rm F}^2 \\
        &{\rm subject~to}~\bm A_m \ge \bm 0, \bm 1^{\top}\bm A_m =\bm 1^{\top}, \bm d \ge \bm 0, \bm 1^{\top}\bm d =1,    \nonumber
\end{align}
where 
$\llbracket \bm d, \bm A_m,\bm A_i,\bm A_j\rrbracket$ is shorthand notation for $ \sum_{k=1}^K \bm d(k)  \bm A_m(:,k) \circ \A_i(:,k)\circ \A_j(:,k)$,
and  $\tT_{m,i,j}$ is the empirical version of $\Expect[\widecheck{\bm{y}}_n^{(m)}\circ\widecheck{\bm{y}}_n^{(i)}\circ{\widecheck{\bm{y}}_n^{(j)}}]$.
{The term ``coupled'' is due to the fact that the tensors $\tT_{m,i,j}$ share one or two latent factors.}
{To further regularize the objective function first- and second-order moments were used in \cite{traganitis2018blind}.}
{
The problem in \eqref{eq:coupledtensor} is nontrivial to solve. An \textit{alternating direction method of multipliers} (ADMM)-based optimization algorithm was employed in \cite{traganitis2018blind} to handle this criterion.  
{Another third-order moment-based DS learning approach was introduced in \cite{zhang2014spectral} that employs an orthogonal tensor decomposition via a robust power method.}
Nonetheless, higher-order moments like  $\tT_{m,i,j}$ in general require a significant amount of samples (annotator labels) to be accurately estimated.

To avoid the sample complexity required for estimating third-order moments,
\cite{ibrahim2019crowdsourcing,ibrahim2021crowdsourcing} proposed {\it  using only second-order statistics}, leading to a coupled nonnegative matrix factorization (CNMF) criterion:
\begin{align}\label{eq:cnmf}
        &\underset{{\{\bm A_m\}_{m=1}^M}, \bm d}{\rm{minimize}} \sum_{\substack{m=1\\i > m}}^M{\rm KL}\left(\bm S_{m,i}||\bm A_m\diag(\bm{d})\bm A_i^{\top}\right) \\
        &{\rm subject~to}~\bm A_m \ge \bm 0, \bm 1^{\top}\bm A_m =\bm 1^{\top}, \bm d \ge \bm 0, \bm 1^{\top}\bm d =1,  \nonumber  
\end{align}
where $\bm S_{m,i}$ is the empirical estimate of
$ 
\Expect[\widecheck{\bm{y}}_n^{(m)}
\circ \widecheck{\bm{y}}_n^{(i)}]$, {$\widecheck{\bm{y}}_n^{(m)}$ {was defined in} \eqref{eq:3rdordertensor1}}, and {${\rm KL}$ denotes the Kullback-Leibler (KL) divergence, defined as ${\rm KL} (\bm p||\bm q)= \sum_{i} [\bm p]_i \log \frac{[\bm p]_i}{ [\bm q]_i}$, with $\bm p$ and $\bm q$ denoting discrete probability distributions of the same support.

The moment-based methods in \cite{traganitis2018blind,zhang2014spectral,ibrahim2019crowdsourcing,ibrahim2021crowdsourcing} come with interesting theoretical support, reminiscent of signal processing research on tensor and nonnegative matrix factorization; see also Sec. \ref{sec:perf_characterization}.

    \noindent
    {\bf Bayesian methods.} 
Under the Bayesian paradigm, \cite{BCC_Kim} considered the following joint probability under presumed priors of $\A_m$ and $\bm d$:
    \begin{align}\label{eq:bayes_post}
        {\sf Pr}(\{\bm A_m\}, \bm d, {\cal Y}|\widecheck{\cal Y}) \propto \prod_{n=1}^N \bm d(y_n) \prod_{m=1}^M \bm A_m(\widecheck{y}_n^{(m)},y_n)\cdot 
        {\sf Pr}(\bm d|\bm \nu)\prod_{m=1}^M\sum_{k=1}^K{\sf Pr}(\bm A_m(:,k)|\bm \pi_{k}^{(m)})
    \end{align}
    where $\bm \nu, \bm \pi_{k}^{(m)} \in \mathbb{R}^K,$ $\forall m,k$ are Dirichlet priors for $\bm d$ and $\bm A_m(:,k)$'s, respectively.
    {Compared to the plain-vanilla DS model learning approaches, Bayesian approaches enjoy more succinct models by treating $\A_m$ and $\bm d$ as random quantities. 
    {However, this comes at the cost of computational intractability; }
    {the posterior in \eqref{eq:bayes_post} is often not easy to evaluate, as it involves marginalization of this joint probability distribution over all parameters.  To circumvent this issue, sampling techniques are often adopted.  
    From the posterior in \eqref{eq:bayes_post}, inference of unknown parameters, {i.e.,} $\{\bm A_m\}, \bm d, {\cal Y}$, is performed using the Gibbs sampling technique by iteratively sampling each parameter from its conditional density function \cite{BCC_Kim}.}   
    Other Bayesian approaches include variational inference techniques by approximating the conditional probability densities using belief propagation and mean field assumptions \cite{liu2012variational}.

    \noindent
{\bf Other methods.} {Beyond methods based on the DS model and its extensions, there are also numerous alternative label integration methods that provide interesting insights and simple implementations. Some examples {leveraging optimization theory, information theory, and geometric interpretations of the label integration problem,} are outlined next.}

{The alternating optimization procedure of the EM algorithm, where annotator parameters and fused labels are iteratively estimated, can also be considered in the absence of a probabilistic model. Indeed, \cite{optbased} formulated label integration as the following optimization problem:}
\begin{subequations} \label{eq:optimization}
    \begin{align}
        \label{eq:optbased}
        \underset{\bm w, \bm y}{\rm minimize} & \sum_{m=1}^{M}w_m \sum_{n=1}^{N} {\sf dist}(y_n,\widecheck{y}_n^{(m)}), \\
        {\rm subject~to} & ~R(\bm w) = 1,
    \end{align}
    \end{subequations}
where $\bm w=[w_1,\dots, w_M]^{\top}$ and $\bm y=[y_1,\dots, y_n]^{\top}$ are the vector of annotator {reliabilities}, similar to the confusion parameter in the one-coin model or the weight values in weighted majority voting, and the vector of ground-truth labels, respectively, {and $R(\bm{w})$ is a generalized constraint (referred to as ``regularization term'' in \cite{optbased}) on the annotator reliabilities}.} The distance measure ${\sf dist}(y_n,\widecheck{y}_n^{(m)})$ captures the deviation from the ground-truth and the annotator responses, e.g., $0-1$ loss for the classification case. {The constraint  $R(\bm{w})$ ensures that the annotator reliabilities $w_m$ remain bounded. The optimization problem in \eqref{eq:optimization} is typically solved by alternating minimization.
For the particular choice of $R(\bm{w}) = \sum_{m=1}^{M}\exp(-w_m)$, \cite{optbased} showed that, given $\bm y$, the weights $\bm w$ admit a closed-form solution. Accordingly, given $\bm w$, and with ${\sf dist}(y_n,\widecheck{y}_n^{(m)})$ set as the $0-1$ classification loss, $\bm{y}$ can be obtained via weighted majority voting; see \eqref{eq:weighted_MV}. Furthermore, this formulation is versatile as different distance measures and constraints can be applied to different contexts.  }

{Leveraging information-theoretic principles, \cite{minimax_entropy} introduced a minimax conditional entropy-based label integration algorithm. The main idea of the proposed algorithm is to find the annotator distributions of maximum entropy, while simultaneously minimizing the entropy of the integrated labels.}
Let $\bm{a}_{n}^{(m)}$ denote the PMF of the noisy label $\widecheck{y}_n^{(m)},$ i.e., $[\bm{a}_{n}^{(m)}]_k = {\sf Pr}(\widecheck{y}_n^{(m)}=k)$. Note that this PMF is dependent on the data item index $n$. Then, {the} following minimize-maximize formulation is considered in \cite{minimax_entropy}:
\begin{align*}
    &\min_{\bm{y}}\max_{\{\bm{a}_n^{(m)} \in {\bm \varDelta}_K\}}  - \sum_{m=1}^{M}\sum_{n=1}^{N}\sum_{k=1}^{K}[\bm{a}_{n}^{(m)}]_k \log [\bm{a}_{n}^{(m)}]_k \\
    &{\rm subject~to}~\sum_{m=1}^{M} [\bm{a}_{n}^{(m)}]_k = \sum_{m=1}^{M} \mathbb{I}[\widecheck{y}_n^{(m)} = k],~ \forall n,k, \\ & \sum_{n=1}^{N} \mathbb{I}[y_n = k'][\bm{a}_{n}^{(m)}]_k = \sum_{n=1}^{N}\mathbb{I}[y_n = k']\mathbb{I}[\widecheck{y}_n^{(m)} = k] \forall m,k,k'.
\end{align*}
Here, the parameters $\bm{a}_n^{(m)}$ are learned by maximizing the entropy of  observations. Ground-truth labels $\bm y$ are inferred by minimizing their entropy, ensuring that the observed labels are the ``least random" choice given the ground-truth label. 
The constraints enforce that the learned PMFs $\bm{a}_n^{(m)}$ align with the empirical observations of the annotator responses. Specifically, the first constraint means that the entries of $\bm{a}_n^{(m)}$ match the numbers of votes obtained per class per data item collectively from all annotators; the second constraint ensures that the learned parameters of each annotator $m$ agree with the empirical estimate of confusions derived from all the responses of that annotator.

    {A geometric interpretation of the MV rule was provided in \cite{maxmargin_MV}. Based on this interpretation, their work introduced a maximum margin algorithm, resembling a support vector machine (SVM), to integrate labels. }
    To illustrate their idea, consider an $M$-dimensional indicator vector $\bm{g}_{n,k}$ for each $n \in [N]$ and $k \in [K]$, with entries $\bm{g}_{n,k}(m) = \mathbb{I}[\widecheck{y}_n^{(m)} = k].$ Then, $\bm w^{\top}\bm{g}_{n,k}$ corresponds to the aggregated score of weighted majority voting  with $\bm w= [w_1,\dots, w_M]^{\top}$ denoting the vector of weight values {(see \eqref{eq:weighted_MV})}. Inspired by the notion of maximum margin in multiclass SVMs, the approach in \cite{maxmargin_MV} seeks a hyperplane that separates the point $\bm{g}_{n,y_n}$ from other points $\bm{g}_{n,k},\;\; k \neq y_n$ by the maximum margin. Consequently, the annotator-specific weights and the ground-truth labels are inferred using the following constrained optimization problem:
   \begin{align*}
       &\underset{\bm w, \{y_n\}}{\rm minimize}~ \frac{1}{2}\|\bm w\|^2\\
       {\rm subject~to}~ &\w^{\top}(\bm{g}_{n,y_n}-\bm{g}_{n,k}) \ge \beta \mathbb{I}[y_n\neq k], \forall n,k,
   \end{align*}
   where $\beta$ represents the maximum margin hyperparameter.

{Finally, interested readers are directed to graph-theoretic based approaches for label integration, including}
 the message passing-based algorithm proposed in \cite{karger2014budget}, {and the $k$-nearest neighbor-based algorithm in \cite{jiang2022learning}}.

     \begin{figure*}[t]
    \centering
    \includegraphics[scale=0.45]{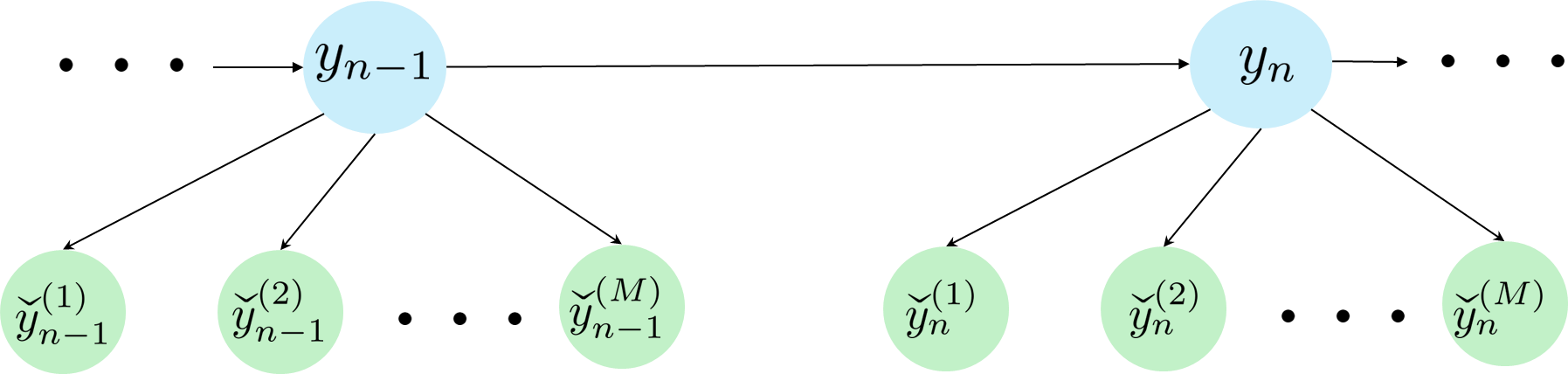}
    \caption{Graphical representation of a hidden {Markov} model for the sequentially dependent data. Green colored circles indicate observed variables, whereas {blue ones indicate latent variables}. 
    }
    \label{fig:hmm_model}
    \end{figure*}

\subsection{Accounting for dependencies}
    \label{ssec:complex}
\noindent
{\bf Dependent data.}
{The models discussed so far consider that data samples are drawn i.i.d. from some unknown distribution. However, structured data often arise, e.g., words in a text or frames in a video.  In these cases,  data samples show {sequential} dependencies, i.e., $\bm x_n$ is dependent on $\bm{x}_{n-1}$ and $\bm{x}_{n+1}$.} Modeling these dependencies can be useful especially in NLP-related crowdsourcing tasks, such as part-of-speech tagging and named-entity recognition, that have recently gained more popularity with the advent of large language models.

{
    {To deal} with sequential data, the DS model can be extended using a hidden Markov model (HMM) \cite{traganitis2020unsupervised}. A simple extension involves a one-step, time-homogeneous {Markov structure to model} the sequence of labels $y_1, y_2, \ldots, y_n$ such that {the} $n$-th label depends only on its immediate predecessor, {i.e.,} ${\sf Pr}(y_n | y_{n-1},\ldots,y_{1}) = {\sf Pr}(y_n|y_{n-1})$---also see Fig.~\ref{fig:hmm_model}. In addition to annotator confusion matrices $\bm A_m$'s and the prior probability vector $\bm d$, the model is also characterized by a $K\times K$ transition matrix $\bm T$ that describes the transitions between labels, i.e., $\bm T(k,k')= {\sf Pr}(y_n=k| y_{n-1}=k')$, for all $n$. Under this DS-HMM model, the joint probability of the observed crowdsourced labels $\widecheck{\cal Y}=\{\widecheck{y}_n^{(m)}\}$ is given by 
    \begin{equation}
    {\sf Pr}(\widecheck{\mathcal{Y}}) = \sum_{\bm k}  \bm d(k_1)\prod_{n=2}^N \bm T(k_n,k_{n-1})\prod_{m=1}^M \bm A_m(\widecheck{y}_n^{(m)},k_n),
\end{equation}
where $\bm k = [k_1,\ldots,k_N]^{\top}\in[K]^N$.

To estimate model parameters, an EM algorithm, similar to the one outlined in Sec.~\ref{ssec:methods_label_int}, can be derived \cite{traganitis2020unsupervised}. The key difference is that the algorithm incorporates a forward-backward algorithm  in the E-step due to the causal nature of the ground-truth labels. The EM algorithm can also be initialized with the solutions obtained from {the moment-based methods discussed in Sec. \ref{ssec:methods_label_int}---see \eqref{eq:coupledtensor} and \eqref{eq:cnmf}}. Here, 
  moments of  annotator responses are characterized {using} the transition probabilities as well, e.g., second-order moments are given by 
\begin{align}\label{eq:3rdordertensor}
\Expect[\widecheck{\bm{y}}_n^{(m)} \circ
        \widecheck{\bm{y}}_n^{(i)}] &= \bm A_m\bm T\diag(\bm{d})\bm A_i^{\top}, \forall m \neq i.
\end{align}
Once the parameters are estimated, a MAP estimate of the ground-truth labels $\mathcal{Y}$ can be obtained via the Viterbi algorithm {\cite{viterbi1967}}.

A Bayesian alternative is introduced in \cite{simpson-seq-2019} that characterizes annotator {confusions} as follows:
\begin{align*}
    \bm C_m(k,k',k'') = {\sf Pr}(\widecheck{y}_n=k|\widecheck{y}_{n-1}=k', y_n={k''}),
\end{align*}
where $\bm C_m$ is a $K \times K \times K$-sized tensor that incorporates the label dependencies as well for the annotator confusions. Under  Dirichlet prior assumptions on the model parameters, the approach maximizes the posterior probability and adopts a variational Bayes-based algorithm for inference. 

Empirical studies of these approaches \cite{traganitis2020unsupervised,simpson-seq-2019} show that considering the dependencies of the data is beneficial and the proposed algorithms are {found to be more promising than those designed for i.i.d. data, in most  scenarios}---{see Table \ref{tab:seq_results} at the end of Sec.~\ref{sec:label_integration}}.
Alternative dependency structures can also be considered, e.g., networked or graph data using a dependency graph that encodes the pairwise relations is handled in \cite{traganitis2020unsupervised}.  

}

            \begin{figure*}[t]
    \centering
    \includegraphics[scale=0.42]{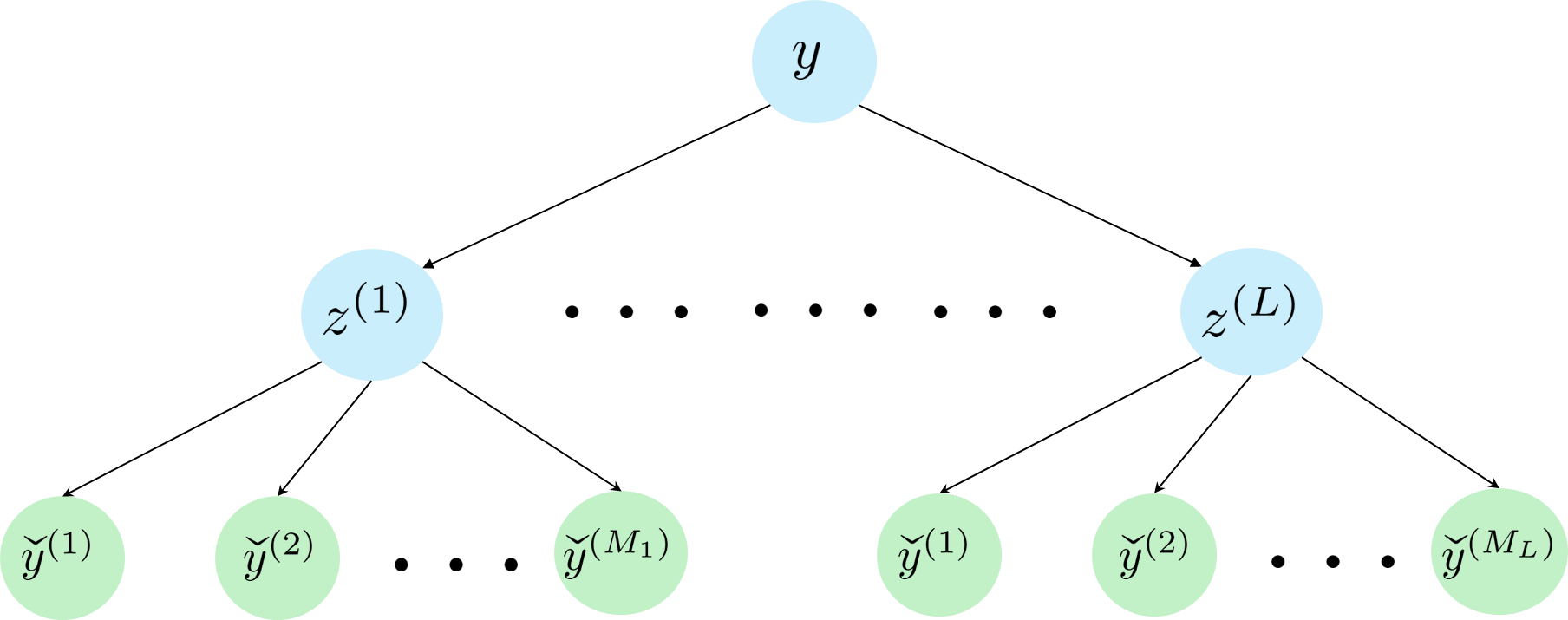}
    \caption{Graphical representation of an extended DS model with $L$ annotator groups. Green colored circles indicate observed variables, whereas blue ones indicate latent variables. 
    }
    \label{fig:ds_hierarchy}
    \end{figure*}

    \noindent
    {\bf Dependent annotators.}
    Recall the key assumption of the DS model, i.e., the annotator responses are conditionally independent. 
    {This is a relatively strong assumption that  may not hold in some cases.} For instance, 
     annotators who underwent similar training may respond similarly  to the same tasks or  spatially close sensors  observing the same phenomenon may capture correlated measurements. As conditional independence no longer holds in these cases, previously introduced methods and algorithms may yield sub-optimal outcomes, due to model misspecification. 
    
    Extending ideas from \textit{distributed detection}, \cite{jaffe2016dependent} introduced a variant to the DS model where dependencies are captured by assigning highly correlated or dependent annotators into {groups}. Assuming that there are $L$ such groups,  annotator responses in the $\ell$-th group are conditionally independent given a latent variable $z^{(\ell)}\in[K]$.
    In addition,  $z_n^{(1)},\dots, z_n^{(L)}$ are conditionally independent given the ground-truth label $y_n$, yielding a hierarchy of DS models---see Fig.~\ref{fig:ds_hierarchy}. {The latent variable $z_n^{(\ell)}$ encodes the ``{collective confusion}'' of the $\ell$-th group, captured in {the} confusion matrix $\bm{\Xi}^{(\ell)}$ such that ${\bm \Xi}^{(\ell)}(k',k)=\prob(z_n^{(\ell)} = k ' | y_n = k).$ Then,} similar to the original DS model, annotator behavior is characterized by confusion matrices w.r.t. different groups, i.e., if an annotator $m$ belongs to group $\ell$, then its confusion matrix $\tilde{\bm A}_m$ is defined as $\tilde{\bm A}_m(k',k)=\prob(\widecheck{y}_n^{(m)} = k' | z_n^{(\ell)} = k).$  
    {A similar model was considered in \cite{vempaty2014reliable}.}

    Under the described model,  \cite{jaffe2016dependent,traganitis2022identifying} proposed to estimate $\tilde{\bm A}_m$'s and $\bm \Xi^{(\ell)}$'s using a hierarchical algorithm. {The {first} step is to estimate the annotators' group membership}. In the case of binary classification, the approach in \cite{jaffe2016dependent} used spectral clustering onto the cross-correlation matrix between different annotator responses in order to assign the annotators to $L$ different groups. This approach was later extended for $K>2$ classes in \cite{traganitis2022identifying}. 
    {Once the annotator's group membership is estimated, the {next} step can employ any DS model learning algorithm within each group $\ell$ to estimate the latent observations $\{z_n^{(\ell)}\}_{n=1}^N$. Using the estimated latent variables $z_n^{(\ell)}$'s, the unknown ground-truth labels $\{y_n\}$'s can be estimated.}

    When annotator dependencies are present, the aforementioned methods show noticeable performance gains compared to methods that are agnostic of said dependencies{---see Fig.~\ref{fig:dependent_annos_results} at the end of Sec.~\ref{sec:label_integration}.}
    {Similar to dependent data, alternative dependency structures between annotators, encoded in a \emph{known} graph, were considered in \cite{BCC_Kim}.}

\subsection{Performance analysis} \label{sec:perf_characterization}

{Quantifying the performance of label integration methods is critical for understanding their effectiveness and limitations.}
A key metric that {captures} the deviation of the corrected labels $\widehat{y}_n$'s from the ground-truth ones $y_n$'s is given by the probability of error (error rate) given by
\begin{equation}
        P_e = \prob(\widehat{y}_n \neq y_n).
\end{equation} 
{Under the DS model, it has been established that}
the error rate $P_e$ of the MAP rule in \eqref{eq:label_integration_DS} decreases exponentially as the number of annotators $M$ increases, i.e.,
\begin{equation}
        P_e \leq \alpha\exp(-\beta M),
\end{equation}
for some constants $\alpha >0 , \beta >0$ \cite{traganitis2018blind,exact_exponent_crowds}.
Similar exponential decrease w.r.t. increasing $M$ was {derived} for the one-coin, confusion vector models as well as for the majority voting rule 
\cite{exact_exponent_crowds}.
These theoretical results {emphasize} the importance of ``crowd wisdom'', {indicating that even basic label integration methods improve with greater $M$.}
Importantly, these results only hold when model parameters are known.
This underscores the significance of having accurate parameter estimation. 
In this subsection, we review some notable theoretical advancements in label integration.

\noindent
{\bf Identifiability of noise models.} 
If the joint PMF ${\sf Pr}(\widecheck{y}_n^{(1)},\ldots,\widecheck{y}_n^{(M)})$ is available, the identifiability of $\bm A_1,\ldots,\bm A_M$ and $\bm d$ under the DS model trivially holds.
This is because the naive Bayes model is also a CPD model with order $M$ and rank $K$. To be more specific, the DS model in \eqref{eq:DSmodel} can be re-written as
\begin{align}\label{eq:rankktensor}
       \underline{\bm P} = \sum_{k=1}^K \bm d(k)  \bm A_1(:,k) \circ \ldots \circ \bm A_M(:,k), 
\end{align}
where $\underline{\bm P}(i_1,\ldots,i_K)={\sf Pr}(\widecheck{y}_n^{(1)}=i_1,\ldots,\widecheck{y}_n^{(M)}=i_K)$, which is an $M$-th order rank-$K$ tensor. The essential uniqueness of $\bm d$ and $\{\bm A_m\}$ in this tensor model holds under mild conditions; see \cite{kargas2017tensors} and the inserted box ``Identifiability of CPD and NMF''.
{Nonetheless, directly estimating ${\sf Pr}(\widecheck{y}_n^{(1)},\ldots,\widecheck{y}_n^{(M)})$ to a reasonable accuracy requires $\Omega(K^M)$ co-labeled samples by all annotators, which can be unrealistic even for moderate $K$ and $M$.}

To circumvent this ``curse of dimensionality'', the identifiability of the DS model using joint distributions of three annotators was established in \cite{zhang2014spectral,traganitis2018blind}. These smaller joint distributions are much more realistic to estimate in practice.
In particular, \cite{traganitis2018blind} showed that the third-order moment term in \eqref{eq:3rdordertensor} is also a rank-$K$ tensor under the CPD model.
Therefore, the optimal solutions to \eqref{eq:coupledtensor} reveal the ground-truth
$\bm A_m, \forall m$ up to a unified permutation ambiguity under mild conditions, e.g., ${\rm rank}(\bm A_m)=K$ for all $m\in [M]$ as ${\rank}(\bm A_m) = {\rm krank}(\bm A_m)$ in this case (see \cite{sidiripoulos2017tensor} and the inserted box ``{Identfiability} of CPD and NMF''). Similar results were derived in \cite{zhang2014spectral} using a different moment construction.
These results are significant, meaning that if at least three annotators co-label a sufficient amount of data, so that the third-order moments can be reliably estimated, {and their confusion matrices are non-singular,} then the DS model is {identifiable}. 
{ 
Note that while the unified permutation ambiguity on the estimated $\widehat{\bm A}_m$'s does not affect reconstructing ${\sf Pr}(X_1,\ldots,X_M)$, it does affect the MAP estimator of \eqref{eq:gt_pedictor}. 
}
{This ambiguity can be resolved {using additional prior knowledge; e.g., by assuming that each annotator has a diagonal dominant $\bm A_m$ \cite{zhang2014spectral,traganitis2018blind}. }}

Identifiability of the DS model using only second-order moments, that is, joint distributions of two annotators' outputs, was established in \cite{ibrahim2019crowdsourcing}.
To be specific, consider a case where seek $\bm A_m$ and $\bm d$ such that $\bm S_{m,i}=\bm A_m{\rm diag}(\bm d)\bm A_i^{\T}$, for $m\in {\cal M}$ and $i\in {\cal I}$, where ${\cal M} \cap {\cal I}=\emptyset$. Suppose ${\cal M}\cup {\cal I}=[M]$.
Then, this is equivalent to 
\begin{align}   
\underbrace{\begin{bmatrix}
    \bm S_{m_1,i_1}, &\ldots,  &\bm S_{m_1,i_T}\\
       \vdots , & \vdots , & \vdots\\
    \bm S_{m_Q,i_1}, &\ldots,  &\bm S_{m_Q,i_T}    
\end{bmatrix} }_{\bm X}=  \underbrace{\begin{bmatrix}
    \bm A_{m_1} \\ \vdots \\ \bm A_{m_Q} 
\end{bmatrix} }_{\bm W} \underbrace{{\rm diag}(\bm d) [\bm A^{\top}_{i_1},\ldots,{\bm A}^{\top}_{i_T}]}_{\bm H}, 
\end{align}
where $Q+T=M$.
Fitting the above model using a KL divergence loss leads to a CNMF formulation similar to \eqref{eq:cnmf}.
As both $\bm W$ and $\bm H$ are nonnegative, the identifiability of these factors holds if both $\W$ and $\H$ satisfy the separability condition from the NMF literature \cite{fu2018nonnegative} --- see the inserted box ``Identifiability of CPD and NMF''. The separability condition holds if $K$ rows of ${\bm W}$ are close to all $K$ unit vectors. This means that there exist (mutually non-exclusive) annotators $m_1,\ldots,m_K$ such that 
$${\sf Pr}(\widecheck{y}_n^{(m_k)}=k|y_n=k)\approx 1 \Longrightarrow \bm A_{m_k}(k,:)\approx \bm e_k;$$ i.e., annotator $m_k$ is an expert of recognizing items from class $k$, leading to the existence of a unit vector in $\bm W$.
In other words, if there are $K$ annotators, who are experts of class $k=1,\ldots,K$, respectively, then $\bm W$ satisfies separability---and the same argument applies to $\bm H$; see \cite{ibrahim2019crowdsourcing,ibrahim2021crowdsourcing} for more relaxed conditions and more advanced settings (e.g., where not all $\bm S_{m,i}$'s are observed).

Beyond the DS model, identifiability was also studied for other noise models. For the one-coin model, model identifiability was established by using the uniqueness (up to {scaling and sign }ambiguity) of the principal eigenvector of $\mathbb{E}[\widecheck{\bm U}\widecheck{\bm U}^\T]$ \cite{ghosh2011moderates}---see ``Spectral Methods'' and Eq.~\eqref{eq:onecoin-ident} in Sec. \ref{ssec:methods_label_int}. 
{The sign ambiguity can be resolved by prior knowledge of one annotator whose probability of being correct is greater than $1/2$, or by making similar assumption as in the CPD and NMF cases.} 
When dependent data are present, the identifiability of both the confusion matrices and the HMM were established in \cite{traganitis2020unsupervised}, also using tensor-based arguments. 
\begin{mdframed}
[backgroundcolor=gray!10,topline=false,
	rightline=false,
	leftline=false,
	bottomline=false]
{\bf Identifiability of CPD and NMF.}
In the context of crowdsourcing, some classical results from the signal processing literature are particularly relevant:

{\it CPD uniqueness.}
Let us denote the $M$-th order rank-$K$ tensor in \eqref{eq:rankktensor} using $\underline{\bm P} = \llbracket \bm d, \bm A_1,\ldots,\bm A_M\rrbracket$. Note that the expression in \eqref{eq:rankktensor} is reminiscent of the SVD of matrices. That is, $\bm A_m\in\mathbb{R}^{I_m\times K}$ has normalized columns, and the term $\bm d(k)$ is analogous to the $k$th singular value.
The model in \eqref{eq:rankktensor} is referred to as the {\it canonical polyadic decomposition} (CPD), which is known to be essentially unique under mild conditions. Specifically, for any alternative $\widetilde{\bm d}$ and $\widetilde{\bm A}_m$ for $m=1,\ldots,M$ such that
$\underline{\bm P}= \llbracket \widetilde{\bm d}, \widetilde{\bm A}_1,\ldots,\widetilde{\bm A}_M\rrbracket$,
it must hold that $\bm A_m=\widetilde{\bm A}_m\bm \Pi $ and $\widetilde{\bm d}=\bm \Pi\bm d$, under mild conditions, {where $\bm \Pi$ is a permutation matrix}.
A widely used condition is $$\sum_{m=1}^M {\rm krank}(\bm A_m) \geq 2K + M -1,$$ where ${\rm krank}$ denotes the Kruskal rank; see \cite{sidiripoulos2017tensor}.

\medskip

{\it NMF uniqueness.}
Consider a low-rank nonnegative matrix $\bm X=\bm W\bm H$ where
$\bm W \in \mathbb{R}_+^{M\times K}$ and $\bm H\in \mathbb{R}_+^{K\times N}$. Then, any nonnegative solution satisfying $\bm X=\widehat{\bm W}\widehat{\bm H}$, $\bm W$ and $\bm H$ 
has to have the form $\widehat{\bm W}=\bm W\bm \Pi\bm \Sigma$
and $\widehat{\bm H}=\bm \Sigma^{-1}\bm \Pi^{\T}\bm H$ under reasonable conditions,
where $\bm \Sigma$ is a diagonal matrix and $\bm \Pi$ is a permutation matrix.
The term $\bm \Sigma$ can be removed if the column norm and row norm of $\bm W$ and $\bm H$, respectively, are known.
Assume that there exist index sets $\bm \Lambda_i$ for $i=1,2$ such that $\bm W(\bm \varLambda_1,:)=\bm D_1$ and $\bm H(:,\bm \varLambda_2)=\bm D_2$, where $\bm D_i$ for $i=1,2$ are full rank diagonal matrices; i.e., both $\W$ and $\H$ satisfy the {\it separability} condition.
Then, the NMF model is essentially unique.
More relaxed conditions for NMF uniqueness exist, e.g., that both $\bm W$ and $\bm H$ satisfy the so-called sufficiently scattered condition (SSC).
Readers are referred to \cite{fu2018nonnegative} for a tutorial on NMF identifiability.

\end{mdframed}

\noindent
{\bf Tractability and scalability.} 
Establishing identifiability of the noise model is only the first step towards successful model learning---as identifiability does not guarantee the existence of a tractable or scalable algorithm.
Some criteria, e.g., the CTD and CNMF objectives in \eqref{eq:coupledtensor} and \eqref{eq:cnmf}, respectively,
present NP-hard optimization problems, and are handled by standard non-convex optimization tools. Although empirical results of these algorithms are often acceptable, the quality of the solution is not theoretically guaranteed.

Nonetheless, some progress has been made towards performance-guaranteed algorithm design.
For example, the {eigendecomposition-based methods for the one-coin model \cite{ghosh2011moderates} and the DS model for binary classification \cite{jaffe2015estimating} {(see ``Spectral Methods'' in Sec. \ref{ssec:methods_label_int})} admit tractable algorithms, as eigen problems are solvable in polynomial time}. If the power method is used, the computational complexity is $O({\rm nnz}(\widecheck{\bm U}))$ per iteration, {with ${\rm nnz}(\cdot)$ denoting the number of nonzero elements,} and the algorithm converges at an exponential rate.
For learning the DS model, similar results were shown in \cite{zhang2014spectral}, where the tensor power method was used.

If class experts with ${\sf Pr}(\widecheck{y}_n^{(m)}=k|y_n=k)\approx 1$ exist for every class, \cite{ibrahim2019crowdsourcing} showed that there is a Gram-Schmidt-like NMF algorithm that recovers $\widehat{\bm A}_m\approx {\bm A}_m\bm \Pi $. This second-order moment matching-based algorithm is also scalable, with a per-iteration complexity of at most $O(MK^3)$, where $K$ is often small.

There are also a number of algorithms that are ``locally convergent'' to the parameters of the DS model. Given sufficiently good initialization, \cite{zhang2014spectral} established that the EM algorithm improves the solution towards the ground-truth confusion matrices. {
There, the tensor power iterations are combined with the scalable EM algorithm (which requires $O(NMK)$ flops per iteration) to provide an overall tractable solution.} 
A similar result for a variational inference algorithm was shown in \cite{traganitis_bayesian}. 
Again if reasonably initialized, \cite{ibrahim2021crowdsourcing} showed that a symmetric NMF algorithm that leverages lightweight Procrustes projection converges exponentially to the ground-truth DS model parameters. {Note that the presence of a few ground-truth labels can reduce the errors of confusion matrix estimates for the EM algorithm~\cite{zhang2014spectral}.}

\begin{table*}[t]
	\centering
	\caption{{Classification Error} ($\%$) and Run-time (sec): AMT Datasets; Table is from \cite{ibrahim2019crowdsourcing}. {The top two results are highlighted in bold.}}
	
	\resizebox{.9\linewidth}{!}{
		\begin{tabular}{l|c|c|c|c|c|c}
			\hline
   \hline
   \textbf{Algorithms} & \multicolumn{2}{c|}{\makecell{\textbf{TREC} \\  ($N=19033$, $M=762$, $K=2$) }} & \multicolumn{2}{c|}{\makecell{\textbf{Bluebird} \\  ($N=108$, $M=39$, $K=2$) }} & \multicolumn{2}{c}{\makecell{\textbf{RTE} \\  ($N=800$, $M=164$, $K=2$) }} \\
   \hline
			& \textbf{(\%) Error} & \textbf{(sec) Time} & \textbf{(\%) Error} & \textbf{(sec) Time} & \textbf{(\%) Error} & \textbf{(sec) Time} \\
			\hline
   \hline
			{CNMF-SPA} \cite{ibrahim2019crowdsourcing}         & 31.47 & 50.68 & 13.88 & 0.07  & 8.75  & 0.28   \\
			\hline
			{CNMF-OPT} \cite{ibrahim2019crowdsourcing} & \textbf{29.23} & 536.89 & \textbf{11.11} & 1.94  & \textbf{7.12} & 17.06  \\
			\hline
			{CNMF-EM} \cite{ibrahim2019crowdsourcing}     & 29.84 & 53.14 & 12.03 & 0.09  & \textbf{7.12} & 0.32   \\
			\hline
			{Spectral-EM} \cite{zhang2014spectral}      & \textbf{29.58} & 919.98 & 12.03 & 1.97  & \textbf{7.12} & 6.40  \\
			\hline
			{CTD} \cite{traganitis2018blind}& N/A      &   N/A    & 12.03 & 2.74  & N/A     &   N/A    \\
			\hline
			{DS-EM} \cite{dawid1979maximum} & 30.02 & 3.20  & 12.03 & 0.02  & 7.25  & 0.07  \\
			\hline
			{Minimax-Entropy} \cite{zhou2014aggregating}   & 30.89 & 352.36 & \textbf{8.33} & 3.43  & 7.50  & 9.10   \\
			\hline
			{SparseSpectralSVD} \cite{dalvi2013aggregating}& 43.95 & 1.48  & 27.77 & 0.02  & 9.01  & 0.03   \\
			\hline
			{KOS} \cite{karger2014budget}  & 51.95 & 9.98  & \textbf{11.11} & 0.01  & 39.75 & 0.03   \\
			\hline
			{SpectralSVD} \cite{ghosh2011moderates} & 43.03 & 11.62 & 27.77 & 0.01  & 49.12 & 0.03  \\
			\hline
			{MV}   & 34.85 &   N/A   & 21.29 &   N/A   & 10.31 &  N/A     \\
			\hline
   \hline
		\end{tabular}%
	}
	\label{tab:amazon}%
\end{table*}% 

\begin{table*}[t]
	\centering
\caption{Real-data results for sequential data. Table from \cite{traganitis2020unsupervised}. {The asterisk $*$ indicates that results are for a subset of available data. {The best results are highlighted in bold.} }}
	\label{tab:seq_results}
 \resizebox{.99\linewidth}{!}{
	\begin{tabular}{ | c | c | c | c || c | c | c | c | c | c | c |}
		\hline
		Dataset & K & M & N & Metric & \textcolor{black}{ Single best} & MV & DS-EM & SeqMM & SeqMM + EM & MV + EM   \\ [0.5ex] \hline
		\multirow{3}{*}{POS} & \multirow{3}{*}{$12$} & \multirow{3}{*}{$10$} & \multirow{3}{*}{$100,676$} &
		  Precision & \textcolor{black}{$0.23$} & $0.22$ & $0.23$ & $0.24$ & $\bf0.25$ & $0.23$ \\\cline{5-11}
		 & & & & Recall & \textcolor{black}{$0.25$} & $0.23$ & $0.25$ & $0.24$ & $\bf0.26$ & $0.24$ \\\cline{5-11}
		 & & & & F1-score & \textcolor{black}{$0.23$} & $0.22$ & $0.22$ & $0.23$ & $\bf0.24$ &  $0.23$ \\ [0.5ex] \hline
		 \multirow{3}{*}{ NER} & \multirow{3}{*}{$9$} & \multirow{3}{*}{$47$} & \multirow{3}{*}{$78,107$} &
		 Precision & { $0.90*$ } & $\bf0.79$ & $0.77$ & $0.74$ & $0.77$ & $0.75$\\ \cline{5-11}
		 & & & & Recall & {$0.24*$ } & $0.59$ & $0.66$ & $\bf0.89$ & $0.69$ & $0.66$ \\ \cline{5-11}
		 & & & & F1-score & {$0.89*$ } & $0.68$ & $0.71$ & $0.62$ & $\bf0.72$ &  $0.70$ \\ \hline
		\multirow{3}{*}{Biomedical IE} & \multirow{3}{*}{$2$} & \multirow{3}{*}{$120$} & \multirow{3}{*}{$7,880,254$} &
		Precision & {$0.94*$} & {$\bf0.89$} & {$0.81$} & {$0.75$} & {$0.69$} & {$0.62$}\\ \cline{5-11}
		& & & & Recall & {$0.76*$} & {$0.45$} & {$0.57$} & {$0.60$} & {$0.68$} & {$\bf0.74$} \\ \cline{5-11}
		& & & & F1-score & {$0.84*$} & {$0.60$} & {$0.66$} & {$0.67$} & {$\bf0.68$} &  {$0.67$} \\ \hline
	\end{tabular}
 }
	\bigskip
\end{table*}

\begin{table*}[t]
\centering\caption{Different label integration approaches and their characteristics}
\resizebox{.99\linewidth}{!}{
\begin{tabular}{c|c|c|c|c|c|c}
\hline
\hline
\textbf{Methods} & \textbf{Model-type} & \textbf{Method-type} & \textbf{Identifiability} & \textbf{Tractabiliy} & \textbf{Scalability} & \textbf{Conditional Independence} \\ 
\hline
\hline
DS-EM \cite{dawid1979maximum}            & DS                  & EM                   &   \xmark                    &      \xmark                &      $\pmb{\checkmark}$               &   $\pmb{\checkmark}$ \\ \hline
CNMF-SPA \cite{ibrahim2019crowdsourcing}             & DS                  & Moment-based Fitting       & $\pmb{\checkmark}$                      &    $\pmb{\checkmark}$                  &  $\pmb{\checkmark}$                    &   $\pmb{\checkmark}$                         \\ \hline
CNMF-OPT \cite{ibrahim2019crowdsourcing}             & DS                  & Moment-based Fitting        & $\pmb{\checkmark}$                      &    \xmark                  &  \xmark                    &                  $\pmb{\checkmark}$          \\ \hline
SymNMF \cite{ibrahim2021crowdsourcing}             & DS                  & Moment-based Fitting        & $\pmb{\checkmark}$                      &    $\pmb{\checkmark}$                 &  $\pmb{\checkmark}$                    &   $\pmb{\checkmark}$                         \\ \hline
CTD  \cite{traganitis2018blind}            & DS                  & Moment-based Fitting       & $\pmb{\checkmark}$                      &             \xmark         &        \xmark              &             $\pmb{\checkmark}$               \\ \hline
Spectral-EM \cite{zhang2014spectral}      &      DS               &   Moment-based Fitting  + EM                    &      $\pmb{\checkmark}$                     &     $\pmb{\checkmark}$                  &       \xmark               &        $\pmb{\checkmark}$                    \\ \hline
Minimax-Entropy \cite{minimax_entropy}  & Extended DS         & Entropy              &  \xmark                        &   \xmark                   &       \xmark                &       $\pmb{\checkmark}$                     \\ \hline
SpectralSVD \cite{ghosh2011moderates}      & One-coin            & SVD                  &     $\pmb{\checkmark}$                     &    $\pmb{\checkmark}$                  &            $\pmb{\checkmark}$          &   $\pmb{\checkmark}$                         \\ 
\hline
SparseSpectralSVD      & One-coin            & SVD                  &          $\pmb{\checkmark}$                &     $\pmb{\checkmark}$                 &    $\pmb{\checkmark}$                  &   $\pmb{\checkmark}$                         \\ 
\hline
GLAD \cite{GLAD}      & Extended One-coin           & EM                  &    \xmark                      &      \xmark                &   $\pmb{\checkmark}$                    &    $\pmb{\checkmark}$                         \\ 
\hline
KOS \cite{karger2014budget}    & Spammer-hammer            & Graph-based                  &           \xmark                &      $\pmb{\checkmark}$                &         $\pmb{\checkmark}$             &    $\pmb{\checkmark}$                        \\ 
\hline
IBCC \cite{BCC_Kim}      & Bayesian-DS            & MCMC                  &     \xmark                     &      \xmark                &                \xmark      &              $\pmb{\checkmark}$              \\ 
\hline
OPT-Crowd \cite{optbased}      & Weighted-MV            & Optimization                  &        \xmark                   &    $\pmb{\checkmark}$                   &             $\pmb{\checkmark}$          &       \xmark                     \\ 
\hline
MaxMarginMV \cite{maxmargin_MV}      & Weighted-MV            & SVM                  &     \xmark                     &    \xmark                  &            $\pmb{\checkmark}$          &       \xmark                     \\ 
\hline
VariationalBayes \cite{liu2012variational}     & Spammer-hammer           & Variational Inference                  &     \xmark                      &     \xmark                   &          \xmark             &         $\pmb{\checkmark}$                    \\ 
\hline
SeqEM \cite{traganitis2020unsupervised}     & DS-HMM            & Moment-based Fitting + EM                  &            \xmark              &     \xmark                 &  \xmark                    &         $\pmb{\checkmark}$                   \\ 
\hline
BayesSeq  \cite{simpson-seq-2019}    & Extended DS-HMM            & Variational Inference                 &        \xmark                  &     \xmark                 &                  \xmark     &                $\pmb{\checkmark}$            \\ 
\hline
GroupAware-EM  \cite{traganitis2022identifying}    & Hierarchical-DS            & Spectral Clustering + EM                 &        \xmark                  &     \xmark                 &                  $\pmb{\checkmark}$    &                \xmark            \\ 
\hline
\hline
\end{tabular}
}

\label{tab:methods}
\end{table*}
\medskip
{ 
\noindent
{\bf Takeaways.}
We use some numerical evidence to conclude our discussion of this section. 
Table~\ref{tab:amazon} shows the performance of a series of label integration methods on real datasets annotated by AMT workers. The annotations are fairly noisy. One can see that majority voting could only correct the labels up to 34.85\%, 21.29\% and 10.31\% error rates for the Text Retrieval Conference (TREC),
Bluebird, and Recognizing Textual Entailment (RTE) datasets, respectively.
However, methods that learn annotator parameters, namely, {DS-EM} \cite{dawid1979maximum}, {Spectral-EM} \cite{zhang2014spectral}, {CNMF} \cite{ibrahim2019crowdsourcing}, and {CTD} \cite{traganitis2018blind} methods, all improve upon the result of MV by large margins.
As discussed, the EM algorithm enjoys low computational complexity.
If it is initialized by a reliable algorithm, e.g., Spectral-EM and CNMF-EM, the EM algorithm often outshines other methods in both accuracy and speed.
These results also attest to the importance of identifiability guarantees---the CNMF and CTD methods learn more accurate labels relative to other methods.

{Table \ref{tab:seq_results} shows results {under} the sequential data setting. In the table, the methods {SeqMM}, {SeqMM + EM}, and {MV + EM} consider data dependencies {in their models}. Specifically, {SeqMM} denotes a moment matching method designed for the DS-HMM model, while {SeqMM + EM} denotes an EM algorithm tailored to the DS-HMM model and initialized with {SeqMM}. {MV + EM} denotes the DS-HMM-based EM algorithm initialized with {MV}.} {Figures of merit include Precision per class, which for class $k$ is defined as $\sum_n \mathbb{I}[\widehat{y}_n = y_n = k]/\sum_n\mathbb{I}[\widehat{y}_n = k]$, and; Recall per class, which is the true positive rate of a classifier, i.e., $\sum_n \mathbb{I}[\widehat{y}_n = y_n = k]/\sum_n\mathbb{I}[y_n = k]$. Precision and recall values reported are averaged across all $K$ classes, alongside their harmonic mean, known as F1-score. In general, these metrics provide useful information for datasets where classes are imbalanced. High precision indicates that a classifier is being conservative in labeling only the data items that it is sure of. High recall indicates that a classifier is prone to multiple false positives. High F1-score indicates a balance between the two, and is generally desirable.}

\begin{figure}[t]
        \centering
                \includegraphics[width=.4\columnwidth]{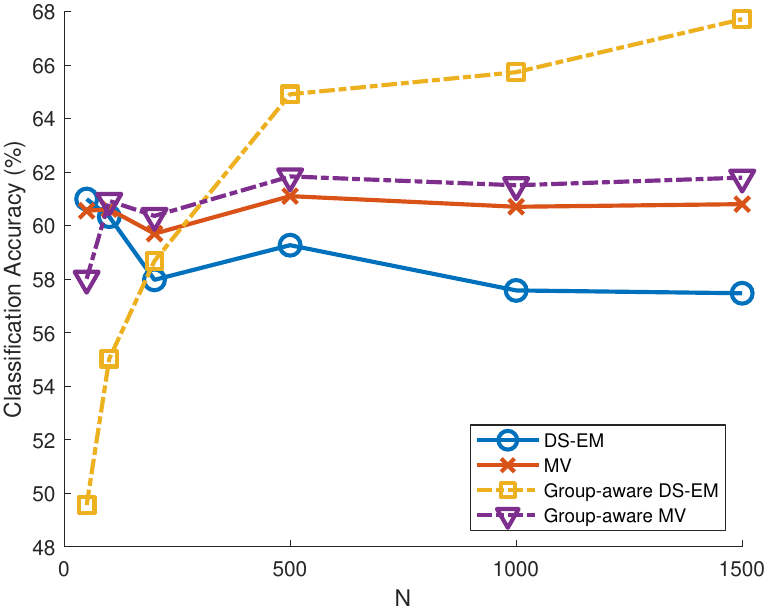}
        \label{fig:classification_acc}
        \caption{Simulated tests on a synthetic dataset with $K=3$
          classes, $M=80$ annotators and $L=4$ groups. ``Group-aware'' algorithms denote algorithms that estimate annotator groups prior to label integration.}\label{fig:dependent_annos_results}
\end{figure}

Fig. \ref{fig:dependent_annos_results} presents the results for the dependent annotator scenario, illustrating the classification accuracies of various methods in a synthetic experiment across different values of $N$. MV and the DS-EM algorithm \cite{dawid1979maximum} are compared against their counterparts that estimate annotator groupings prior to DS model estimation, denoted as {Group-aware MV} and {Group-aware DS-EM}, respectively. From the results, one can note that the ``group-aware'' algorithms outperform their group-agnostic counterparts.}

Table \ref{tab:methods} presents a summary of the several methods discussed in this section along with their characteristics. The {\it key takeaways} are as follows: First, identifiability of noise models is often a key performance indicator. Empirical evidence strongly suggests that  approaches with identifiability guarantees consistently work well for label integration. 
Second, sample (annotator label) complexity is a key consideration when selecting an algorithm in practice. Since collecting more labels incurs extra costs and increases annotator workload, methods that perform well with fewer annotations are more economical. 
Third, methods need to be scalable as real-world label integration often involves a large number of data items and annotators.

\section{End-to-End (E2E) Learning from 
    Crowdsourced Labels}  \label{sec:e2e}

Compared to the label integration paradigm,
the E2E one depicted in Fig.~\ref{fig:e2e} has shown more appealing performance over various datasets (cf. Fig.~\ref{fig:2vse2e}). This is because E2E {crowdsourcing} incorporates features to jointly tackle label integration and downstream learning, and thus alleviates error accumulation and propagation.

\begin{figure*}[t!]
    \centering
    \includegraphics[scale=0.4]{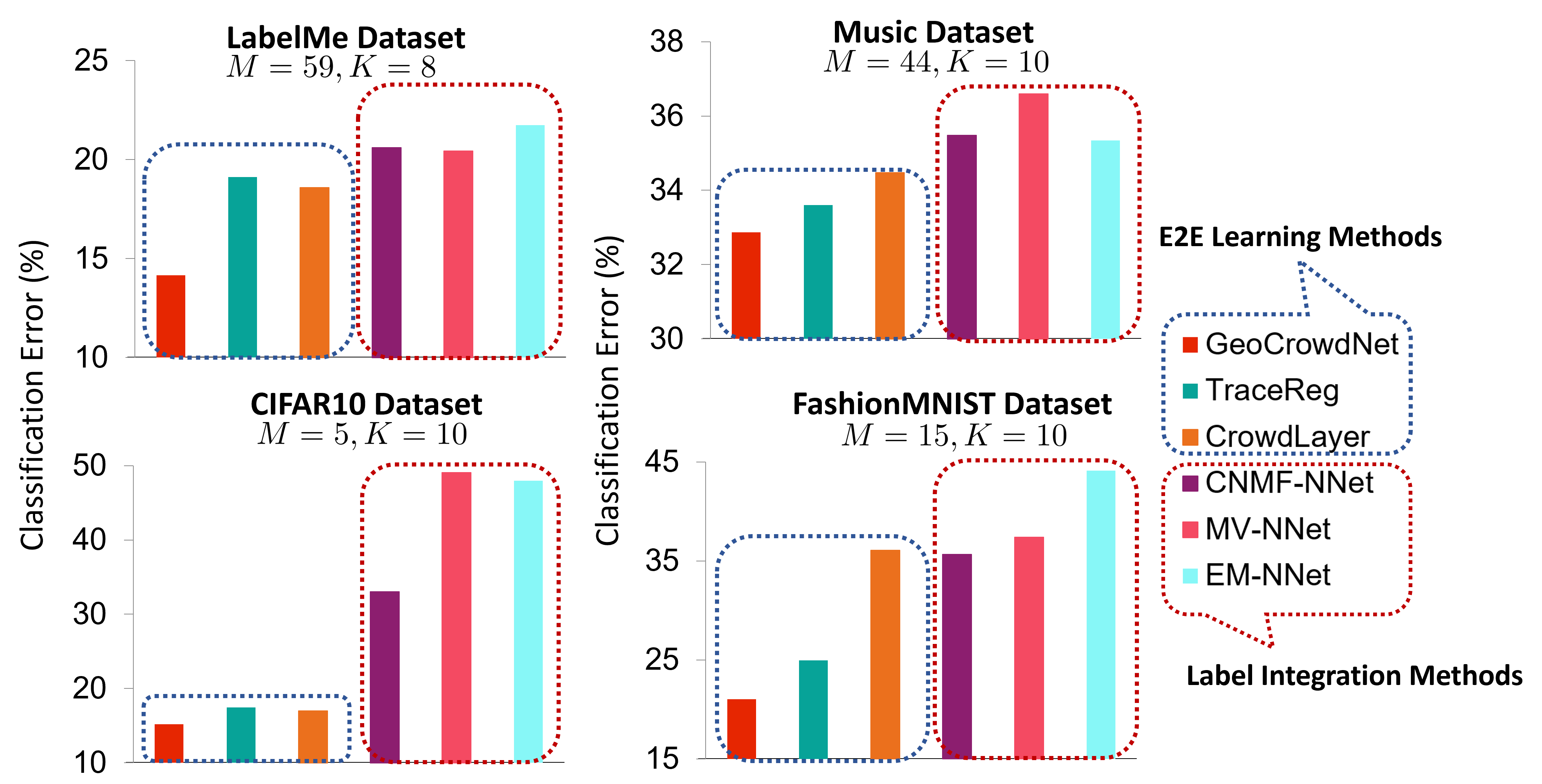}
    \caption{The performance of E2E methods, GeoCrowdNet \cite{ibrahim2023deep}, TraceReg \cite{tanno2019learning}, and CrowdLayer \cite{rodrigues2018deep} and the {two-stage} approaches (label {integration} methods CNMF \cite{ibrahim2021crowdsourcing}, MV, and DS-EM \cite{dawid1979maximum} followed by training a neural network classifier. {The top row presents experiments run on LabelMe \cite{russell2007labelme} and Music \cite{rodrigues2014gaussian} datasets, where  labels are provided by human workers from AMT. The bottom row presents  synthetic experiments run on CIFAR-10 and FashionMNIST datasets, where confusion matrices $\bm A_m$'s are synthetically generated. Results are taken from \cite{ibrahim2023deep} and illustrated to highlight the performance gap between label integration methods and E2E learning methods. 
    }}
    \label{fig:2vse2e}
\end{figure*}

\subsection{E2E learning via maximum likelihood and EM}
With ${\cal D} := \{\bm x_n, \{\widecheck{y}^{(m)}_n\}_{m=1}^M \}_{n=1}^N$ collecting independent data, and annotator responses assumed conditionally independent given ground-truth labels, as in the DS model, consider the joint likelihood
    \begin{align} 
        {\sf Pr}({\cal D}) &=  \prod_{n=1}^N {\sf Pr}(\bm x_n, \widecheck{y}_n^{(1)},\dots,\widecheck{y}_n^{(M)}) 
        = \prod_{n=1}^N \sum_{y_n=1}^K  {\sf Pr}(y_n |\bm x_n) \prod_{i=1}^M {\sf Pr}(\widecheck{y}_n^{(m)}|y_n,\x_n) \label{eq:mle2}\\
          &= \prod_{n=1}^N \sum_{y_n=1}^K  \underbrace{ {\sf Pr}(y_n |\bm x_n) }_{[\bm f^\star(\bm x_n)]_k}\prod_{m=1}^M \underbrace{{\sf Pr}(\widecheck{y}_n^{(m)}|y_n)}_{\bm A_m(\widecheck{y}_n^{(m)},y_n)} \label{eq:mle3}
    \end{align}
where we used the conditional independence of 
$\widecheck{y}_n^{(m)}$ given $y_n$, the assumption that annotator confusion is independent from data items, and 
$\bm f^\star$, $\bm A_m$ are defined as in 
\eqref{eq:gt_pedictor} and \eqref{eq:confusionmatrix}. {Again, $\bm A_m$ is assumed invariant across $n$, and conditioned on $y_n$, annotator labels $\check{y}_n^{(m)}$ are also independent across $n$.}

Given $\cal D$, the goal of E2E learning is to find ${\sf Pr}(y_n |\bm x_n)$. With classification being the downstream task, this posterior distribution maps any data item to a PMF over the class labels $1,\ldots,K$. {Let $\bm f_{\bm \theta}:\mathbb{R}^D\rightarrow \mathbb{R}^K$ be a function parameterized by $\bm \theta$ that represents the ground-truth {label posterior}} $\bm f^\star$.
Collecting all model parameters in ${\bm \psi} := ({\bm A}_1,\dots, {\bm A}_M, {\bm \theta} )$, we have $${\sf Pr}({\cal D};\bm \psi)= \prod_{n=1}^N \sum_{y_n=1}^K  [\bm f_{\bm \theta}(\bm x_n)]_{y_n} \prod_{m=1}^M \bm A_m(\widecheck{y}_n^{(m)},y_n).$$
Similar to label integration in Sec. \ref{ssec:methods_label_int}, the MLE yields \begin{align} \label{eq:mle_psi}
        \widehat{\bm \psi}  = \underset{\bm \psi}{\text{arg~max}}~~\log \left({\sf Pr}({\cal D};\bm \psi)\right)
\end{align} 
where maximizing over $\bm \psi$ is typically nontrivial (non concave). As before, this motivates adopting the EM 
\cite{raykar2010learning,rodrigues2018deep}.  Viewing ${\cal Y} := \{y_n\}_{n=1}^N$ as  latent variables, the 
expected value of the complete log-likelihood $\log({\sf Pr}({\cal D},{\cal Y};\bm \psi))$ under the current estimate of $\bm \psi$ is computed in the E-step:
    \begin{align*}
        Q(\bm \psi; \bm \psi^t) &= \mathbb{E}_{{\cal Y}\sim {\sf Pr}({\cal Y};{\cal D}, \bm \psi^t)}[\log {\sf Pr}({\cal D},{\cal Y}; \psi)] = \sum_{n=1}^N \sum_{k=1}^K q(y_n=k; \bm \psi^t)  \log {\sf Pr}(\bm x_n, y_n = k, \widecheck{y}_n^{(1)},\dots,\widecheck{y}_n^{(M)};\bm \psi)
    \end{align*}
    with
    $
        q(y_n=k; \bm \psi^t)  = \frac{1}{Z}  [{\bm f}_{\bm \theta^t}(\bm x_n)]_k \prod_{m=1}^M \bm A_m^t(\widecheck{y}_n^{(m)},k)
    $
    and $Z$ being the normalization constant---see \eqref{eq:DS_EM_posterior}. 

    The M-step estimates $\bm \psi$ by maximizing $Q(\bm \psi; \bm \psi^t)$. This can be done by alternating between the following updates:  
    \begin{subequations}
      \begin{align} \label{eq:Am_updates}
        \bm A^{t+1}_m (k',k) &= \frac{\sum_{n=1}^N q(y_n=k; \bm \psi^t) \mathbb{I}[\widecheck{y}_n^m=k']}{\sum_{k''=1}^K\sum_{n=1}^N q(y_n=k; \bm \psi^t)\mathbb{I}[\widecheck{y}_n^m=k'']}\\
        \bm \theta^{t+1} &=  \arg\max_{\bm \theta} ~Q(\bm \theta; (\bm \theta^{t},\{\bm A_m^{t+1}\})). \label{eq:thetaupdate}
    \end{align}
    \end{subequations}
This EM formulation is very similar to the one {outlined} for the DS model in Sec.~\ref{ssec:methods_label_int}, albeit with the classifier $\bm f_{\bm \theta}$ taking the role of the prior class probabilities $\bm d$.
The EM framework is flexible in terms of incorporating various $\bm f_{\bm \theta}$ function classes. For binary classification,\cite{raykar2010learning} advocated a logistic regression model, where $[\bm f_{\bm \theta}]_1 = \sigma(\bm \theta^{\top}\bm x_n)$ and $[\bm f_{\bm \theta}]_{2} = 1-\sigma(\bm \theta^{\top}\bm x_n)$,
and $\sigma$ denotes the sigmoid function. As a result, a Newton-Raphson algorithm can be implemented for \eqref{eq:thetaupdate}.
Neural networks were used in \cite{rodrigues2018deep} to serve as $\bm f_{\bm \theta}$, where \eqref{eq:thetaupdate} was updated by back-propagation based stochastic gradient.
To deal with sequence-type data, \cite{rodrigues2014sequence} used a conditional random field (CRF) function as the classifier, which utilizes the Viterbi algorithm and the limited-memory Broyden–Fletcher–Goldfarb–Shannon (BFGS) algorithm for the M-step. A Bayesian method was adopted in \cite{rodrigues2014gaussian} that used a Gaussian process to model a binary classifier and adopted an expectation propagation (EP)-based algorithm for the inference that involves EM-like iterative steps.

\subsection{Deep learning with ``Crowd Layer''}
\label{ssec:crowdlayer}
Among all the functions that can be used as $\bm f_{\bm \theta}$, deep neural networks (DNNs) {attract} a lot of attention, due to their remarkable empirical success in various domains.
While \cite{rodrigues2018deep} showed that EM can be used together with DNNs, the EM framework has some limitations.
First, the EM framework is based on multi-class classification, yet it is not straightforward to extend it to cover other problem settings, e.g., when sequence data is involved---{the E-step could quickly become intractable}. 
Second, the derivation of the EM framework relies on the conditional independence of the annotators, which may not be always a valid assumption, as discussed in Sec.~\ref{ssec:complex}.
Third, the function $\bm f_{\bm \theta}$ needs to be trained in each M-step, which may be computationally demanding.

An alternative approach to incorporate DNNs in crowdsourcing was {put forth} in \cite{rodrigues2018deep}.
Consider the {conditional} probability of $m$-th {annotator response} to the data item $\bm x_n$ as follows:
 \begin{align}\label{eq:bestmodel}
	{\sf Pr}(\widecheck{y}^{(m)}_n=k |\bm x_n) = \sum_{k'=1}^K{\sf Pr}(\widecheck{y}^{(m)}_n=k |y_n=k'){\sf Pr}(y_n=k' |\bm x_n),~k \in [K],
	\end{align}
where we used the law of total probability and the assumption that annotator responses are instance-independent, given the label $y_n$.  
Upon defining a $K$-dimensional vector $\bm p_n^{(m)}$ such that $[\bm p_n^{(m)}]_k \triangleq  {\sf Pr}(\widecheck{y}^{(m)}_n=k |\bm x_n)$, Eq.~\eqref{eq:bestmodel} can be expressed as follows:
\begin{align} \label{eq:model}
	{\bm p}_{n}^{(m)} = \bm A_m \bm f^\star(\bm x_n), \forall m,n,
\end{align}
where $[\bm f^\star(\x_n)]_k={\sf Pr}(y_n=k|\x_n)$ is as defined in \eqref{eq:gt_pedictor}.
Under this model, observations can be understood as realizations of a categorical random variable, {i.e.,} $\widecheck{y}^{(m)}_n {|\bm{x}_n }\sim {\rm categorical}({\bm p}_{n}^{(m)})$. 
To estimate $\bm A_m$ and $\bm f^\star$, a commonly used criterion in machine learning is {\it cross entropy} (CE), i.e.,
\[      {\rm CE}(\widecheck{\bm y}_n^{(m)}, \bm A_m \bm f_{\bm \theta}(\bm x_n) ) = -  \sum_{k=1}^K[\widecheck{\bm y}_n^{(m)}]_k\log [\bm A_m \bm f_{\bm \theta}(\bm x_n)]_k,   \]
where {$\widecheck{\bm y}_n^{(m)}$ is the one-hot encoding of $\widecheck{y}_n^{(m)}$, i.e.,} $\widecheck{\bm y}_n^{(m)}(k)=1$ if $\widecheck{y}_n^{(m)}=k$ and $\widecheck{y}_n^{(m)}(k')=0$ for $k'\neq k$, and $\bm f_{\bm \theta}$ is the learning function for approximating $\bm f^\star$ as before. {CE is related to the KL divergence as, for two distributions $\bm{p},\bm{q}$, ${\rm KL} (\bm p||\bm q)= \sum_{i} [\bm p]_i \log \frac{[\bm p]_i}{[\bm q]_i} = \sum_i [\bm{p}]_i\log[\bm{p}]_i + {\rm CE}(\bm p,\bm q)$. Thus, minimizing CE}
seeks a model $\{\bm A_m, \bm f_{\bm \theta}(\bm x_n)\}$ that matches the ``empirical PMF'' $\widecheck{\bm y}_n^{(m)}$. It can be shown that when $N\rightarrow \infty$, the minimum of CE is attained at $\bm A\bm f_{\bm \theta}(\bm x_n) = \bm p^{(m)}_n$. Collecting all annotator responses, \cite{rodrigues2018deep} used the following \textit{coupled cross-entropy minimization criterion} (CCEM):
		\begin{align}\label{eq:unreg}
		 \underset{\bm f_{\bm \theta}\in {\cal F}, \{\bm A_m \in {\cal A}\}_{m=1}^M}{\rm minimize}&~~-\frac{1}{|{\cal S}|}\sum_{(m,n) \in {\cal S}} \sum_{k=1}^K\mathbb{I}[\widecheck{y}_{n}^{(m)}=k]  \log [\A_m \bm f_{\bm \theta}(\x_n)]_k, %\\
		\end{align}
 where ${\cal S}\subseteq [M]\times [N]$ is the index set of annotator-labeled samples, $\mathcal{F} \subseteq \{\bm f(\x) \in\mathbb{R}^K | ~\bm f(\x) \in \bm \varDelta_K,~\forall \x \}$ is a function class parameterized by $\bm \theta$, {$
\bm \varDelta_K$ represents the $(K-1)$-probability simplex,}
and ${\cal A}$  is the constrained set of confusion matrices $\{\bm A \in \mathbb{R}^{K \times K} |\bm A \ge  \bm 0, \bm 1^{\top}\bm A = \bm 1^{\top}\} $.
{In practice, $\bm f_{\bm \theta}\in {\cal F}$ can be approximately enforced by using a softmax layer as its output.}
The constraints ensure that the output of $\bm f_{\bm \theta}$ and columns of $\bm A_m$'s are PMFs.
The term ``coupled'' refers to the fact that the expressions of $\bm p^{(m)}_n$ for all $m$ are coupled by  $\bm f_{\bm \theta}(\x_n)$.

Fig.~\ref{fig:crowdlayer} illustrates the loss function of CCEM. Here, confusion matrices $\bm A_m$'s act as additional annotator-specific layers  of the neural network \cite{rodrigues2018deep}, hence this approach is termed \emph{crowdlayer}.
The CCEM type formulation is arguably more versatile in terms of modeling and computation relative to the EM-type E2E algorithms. First, regularization terms on $\bm A_m$ and $\bm \theta$ can be easily added for various purposes, e.g.,
incorporating prior knowledge and enhancing identifiability \cite{ibrahim2023deep,tanno2019learning}.
Second, as we have seen, the CCEM approach does not require that  annotators are conditionally independent (see details in \cite{ibrahim2023deep}).
{Third, the empirical distribution $\widecheck{\bm y}_{n}^{(m)}$ needs not be categorical, opening doors for continuous measurement-based problems, e.g., regression.} 
{Finally, the crowdlayer architecture can be easily trained via backpropagation and any gradient decent-based optimization algorithms like Adam, using off-the-shelf deep learning libraries such as PyTorch and TensorFlow.}

\begin{figure*}[t]
    \centering
    \includegraphics[scale=0.45]{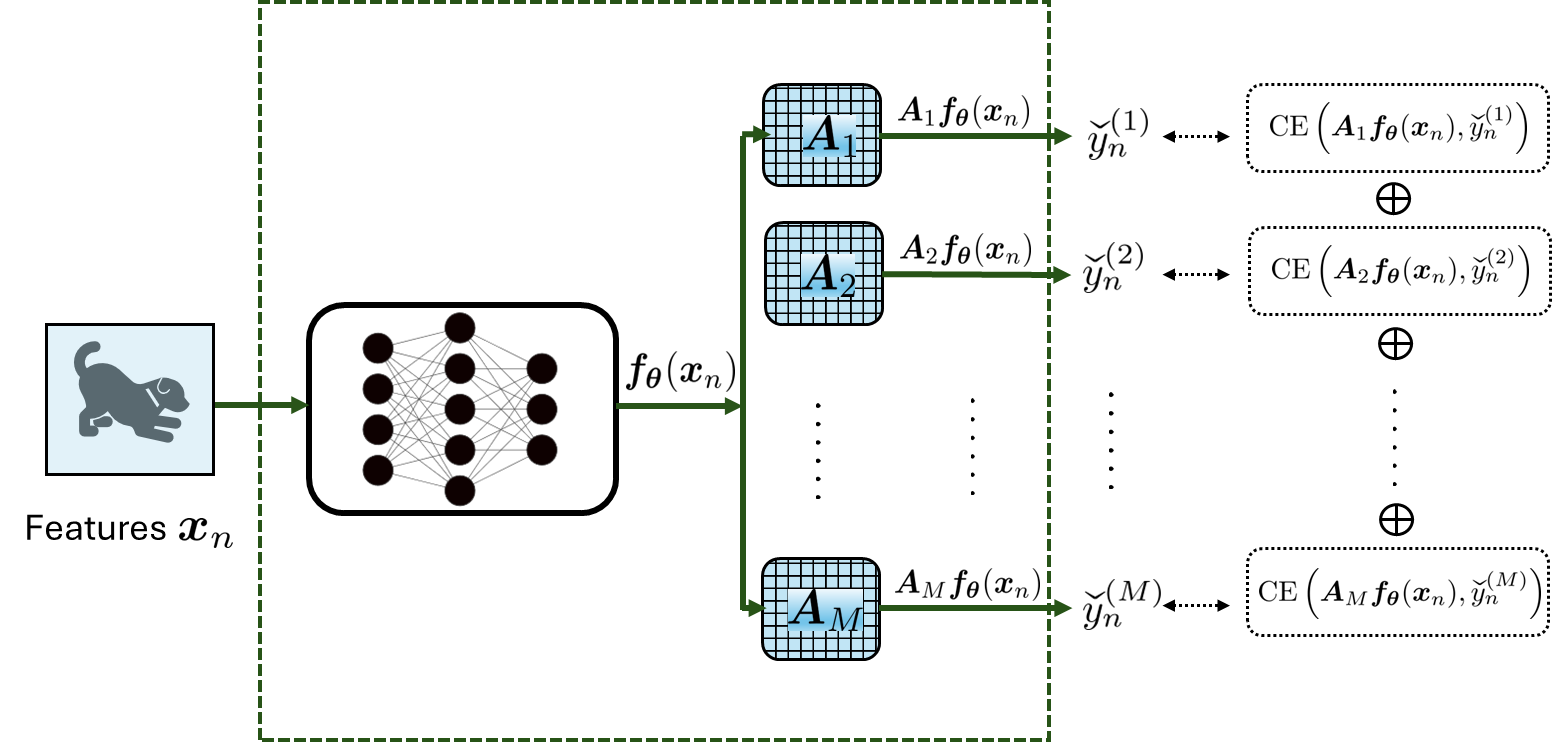}
    \caption{The ``crowdlayer"-based architecture for deep learning-based E2E crowdsourcing { \cite{rodrigues2018deep,ibrahim2023deep}.}
    }
    \label{fig:crowdlayer}
\end{figure*}

\subsection{Model identifiability under CCEM}
Issues of model identifiability also arise in the E2E context. 
Under the CCEM criterion, and the generative model $\widecheck{y}_n^{(m)} {| \bm{x}_n} \sim {\rm categorical}(\A_m\f^\star(\x_n)),$
it is  essential to identify $\bm A_m$'s and the ground-truth $\bm f^\star(\x_n)$ (i.e., ${\sf Pr}(y_n|\x_n)$).
Note that identifying $\bm f^\star(\x_n)$ (i.e., to attain $\bm f_{\bm \theta}(\x_n)\approx \bm f^\star(\x_n)$) over all the seen (training) data $\{\x_n\}_{n=1}^N$ allows the learned model $\bm f_{\bm \theta}$ to generalize over unseen (test) data as well.
This turns out to be non-trivial.
In the ideal case where $N\rightarrow\infty$, 
the CCEM criterion returns $\widehat{\bm f}_{\bm \theta}$ and $\widehat{\bm A}_m$ such that
${\bm p}_{n}^{(m)} = \widehat{\bm A}_m \widehat{\bm f}_{\bm \theta}(\bm x_n)$. Nevertheless, one can easily note that such a relation is highly non-unique since there {may exist many possible} nonsingular matrices $\bm Q \in \mathbb{R}^{K \times K}${, not necessarily limited to being permutation ambiguity matrices,} such that $\bm p_{n}^{(m)} =(\widehat{\bm A}_m \bm Q) (\bm Q^{-1} \widehat{\bm f}_{\bm \theta}(\bm x_n))$. However, CCEM-type approaches seem to always learn reasonable $\bm A_m$ and $\bm f_{\bm \theta}$ in practice.
To understand this phenomenon, \cite{ibrahim2023deep} provided performance characterizations of CCEM, from an NMF identifiability viewpoint.
To be specific, when $N\rightarrow\infty$, the CCEM criterion can be understood as finding solutions to fit the following model:
\begin{align} \label{eq:nmf_1}
	\underbrace{\begin{bmatrix}
		{\bm p}_{1}^{(1)}&\dots& {\bm p}_{N}^{(1)}\\
		\vdots&\ddots&\vdots\\
		{\bm p}_{1}^{(M)}& \dots& {\bm p}_{N}^{(M)}\\
		\end{bmatrix}}_{\bm P \in \mathbb{R}^{MK \times N}} = \underbrace{\begin{bmatrix}\bm A_1\\\vdots\\\bm A_M\end{bmatrix}}_{\bm W \in \mathbb{R}^{MK \times K}} \underbrace{\begin{bmatrix}\bm f^\star(\bm x_1)& \dots &\bm f^\star(\bm x_N) \end{bmatrix}}_{\bm H \in \mathbb{R}^{K \times N}},
	\end{align}
where the factors $\W$ and $\bm H$ are both nonnegative per their {definition}. Clearly, if $\bm W$ and $\bm H$ both satisfy the separability condition (or the sufficiently scattered condition (SSC)), as  discussed in ``Identifiability of CPD and NMF'' (also see \cite{fu2018nonnegative}), the factorization model in \eqref{eq:nmf_1} is essentially unique; i.e.,
there exists a permutation matrix $\bm \Pi$ such that
\begin{align}\label{eq:ident_Af}
  \|\widehat{ \bm A}_m\bm \Pi - \bm A_m\| \rightarrow 0,~~\| \bm \Pi^\T\widehat{\bm f}_{\bm \theta}(\bm x) - \bm f^\star(\bm x)\| \rightarrow 0,
\end{align}
when $N\rightarrow \infty$.
For $\bm W$, the separability condition still requires the existence of class experts for each of the $K$ classes.
For $\bm H$, separability implies the existence of $K$ indices, $\{n_1,\ldots,n_K\}=\{1,\ldots,K\}$ such that
$\bm f^\star(\x_{n_k})=\bm e_k \Leftrightarrow {\sf Pr}(y_{n_k}=k|\x_{n_k})=1$.  Such points $ \x_{n_k}$  are called ``anchor points'' for the classes and are often found useful in noisy label learning.

Aiming to relax the conditions on one of $\bm W$ and $\bm H$, {\cite{ibrahim2023deep} introduced regularization to the CCEM criterion.}
Specifically, a regularization term to maximize the volume of $\bm H=[\bm f_{\bm \theta}(\x_1),\ldots,\bm f_{\bm \theta}(\bm x_N)]$ was added:
\begin{align}
    \minimize_{ \{\bm A_m\in {\cal A}\},\bm f_{\bm \theta}\in \bm \varDelta_K }~{\cal L}_{\rm CCEM} - \beta \log\det(\bm H\bm H^\T),
\end{align}
where ${\cal L}_{\rm CCEM}$ is defined as the objective function in \eqref{eq:unreg} and $\beta\geq 0$.
This formulation leverages the {\it simplex volume minimization}-based structured matrix factorization \cite{fu2018nonnegative} to establish identifiability as shown in \eqref{eq:ident_Af}.
This way, the separability or SSC condition imposed onto $\bm W$ can be removed. 
This substantially reduces the expertise requirement for the annotators to establish identifiability of the model.
s to establish identifiability of the model.

\subsection{{Alternative} E2E crowdsourcing approaches}

\noindent
{\bf Agreement-based model.} 
{Besides the confusion matrix-based models and formulations as in \eqref{eq:mle_psi} and \eqref{eq:unreg}, {alternative treatments have been proposed to enable E2E learning.}
{For example,} the approach in \cite{peng2019maxmig} 
proposes to use an annotator ``aggregator'' denoted as $\bm g_{\bm \phi}$ that measures the ``average agreement'' between different annotator responses, and is modeled as an affine mapping followed by a so-called softmax operation:
\begin{align*}
      \bm g_{\bm \phi}(\{\widecheck{\bm y}_n^{(m)}\}) = {\rm softmax}\left(\sum_{m=1}^M\bm W^{(m)}\widecheck{\bm y}_n^{(m)}+\bm b\right).
\end{align*}
Here, $\bm \phi = \{\bm W^{(1)}, \dots, \bm W^{(M)},\bm b\}$ and $\widecheck{\bm y}_n^{(m)}$ denotes the one-hot embedding of the noisy label $\widecheck{y}_n^{(m)}$ as before. 
{Note that $\bm f_{\bm \theta}(\bm x_n)$ and $\bm g_{\bm \phi}(\{\widecheck{\bm y}_n^{(m)}\}) $ can be understood as two label predictors using data features and annotator-produced labels as inputs, respectively.}
The idea in \cite{peng2019maxmig} is to maximize the ``agreement''  between $\bm f_{\bm \theta}$ and $\bm g_{\bm \phi}${, using an $f$-mutual information gain (${\rm MIG}^f$}) objective}:  
\begin{align}\label{eq:maxmig}
\underset{\bm \theta, \bm \phi}{\rm maximize} ~{\rm MIG}^f(\bm f_{\bm \theta},\bm g_{\bm \phi};\{\bm x_n\},\{\widecheck{\bm y}_n^{(m)}\}  ).
\end{align}
{In essence, ${\rm MIG}^f$ measures the agreement between $\bm f_{\bm \theta}(\bm x_n)$ and $\bm g_{\bm \phi}(\{\widecheck{\bm y}_n^{(m)}\})$ averaged over all the data items. The measure of agreement is an $f$-divergence loss function, such as the KL divergence.
It was also shown in \cite{peng2019maxmig} that the objective \eqref{eq:maxmig} finds optimal solutions (i.e., solutions that extract maximum information from their inputs to predict the ground-truth) in the asymptotic case, provided there are conditionally independent expert annotators.}

\noindent
{\bf Instance-dependent confusion matrix.}
Both EM and CCEM based E2E methods assume that $\bm A_m$ remains identical across all $\x_n$ {(see \eqref{eq:model})}.
Recently, crowdsourcing approaches under an instance-dependent setting have been advocated {\cite{zhang2020disentangling,guo2023label,nguyen2024noisy}} where
\begin{align} \label{eq:insatnce_sig_model}
	{\bm p}_{n}^{(m)} = \bm A_m(\bm x_n) \bm f^\star(\bm x_n), \forall m,n.
\end{align}
Here, $\bm A_m(\bm x_n)$'s are instance-dependent annotator confusions with entries $[\bm A_m(\bm x_n)]_{k,k'}= {\sf Pr}(\widecheck{y}^{(m)}_n=k |y_n=k', \bm x_n)$; {see also \eqref{eq:mle2}}. 
{
To estimate these instance-dependent confusion matrices,
a common approach is to use two learnable functions, i.e., $\bm f_{\bm \theta}:\mathbb{R}^{D}\rightarrow \mathbb{R}^K$ and $\bm A^{\bm \phi}_m(\cdot):\mathbb{R}^{D}\rightarrow\mathbb{R}^{K\times K}$ (e.g., neural networks) to parameterize $\bm f^\star$ and $\bm A_m(\x_n)$, respectively.

{A multi-stage learning strategy is proposed in \cite{guo2023label}}, i.e., first learning $\bm A^{\bm \phi}_m(\cdot)$ using some {pre-selected} data items 
and then using the learned $\bm A^{\bm \phi}_m(\cdot)$ to train $\bm f_{\bm \theta}$ using losses similar to \eqref{eq:trace_instance_dep}. {Specifically},  \cite{guo2023label} parametrizes the instance-dependent annotator confusions using a \textit{mixed effects neural network} model (MNN):
\begin{align}
    [\bm A_m^{\bm \phi}(\bm x_n)]_{:,k} = {\rm softmax}(\underbrace{\bm B^{(m)} \bm g_{\bm \phi_1}(\bm x_n)}_{\text{annotator-specific}} + \underbrace{\bm C^{(k)} \bm g_{\bm \phi_2}(\bm x_n)}_{\text{class-specific}}),
\end{align}
where $\bm g_{\bm \phi_1}$ and $\bm g_{\bm \phi_2}$ are two neural networks parameterized by $\bm \phi_1$ and $\bm\phi_2$, respectively. The model aims to capture the the annotator-specific and class-specific 
effects on $\bm A_m(\x_n)$. In this work, the first step selects some ``anchor data points'' for each class $k$. The anchor points satisfy ${\sf Pr}(y_n=k|\bm x)=1$, which means $[\bm A_m(\bm x_n)]_{:,k}= {\sf Pr}(\widecheck{y}^{(m)}_n=k|\bm x_n)$. Using such selected data items $\{\bm x_s,\{\widecheck{y}_n^{(m)}\} \}_{s=1}^S$,  MNN parameters $\bm \varphi=\{\bm \phi_1, \bm \phi_2, \{\bm B^{(m)}\}, \{\bm C^{(k)}\}\}$ are learned first using a regression approach. 
{The next step involves predicting the ground-truth labels through a pairwise likelihood ratio test using the learned $\bm{A}_m^{\bm{\phi}}(\cdot)$. These predictions are then used to train $\bm{f}_{\bm{\theta}}$.} 

{E2E learning of the model in \eqref{eq:insatnce_sig_model}
was considered in
\cite{zhang2020disentangling}, which extended  \cite{tanno2019learning} as follows: 
\begin{align}\label{eq:trace_instance_dep}
		\minimize_{{\{\bm A_m^{\bm \phi}\in {\cal A}\},\bm f_{\bm \theta}\in \bm \varDelta_K}}&-\frac{1}{|{\cal S}|}\sum_{(m,n) \in {\cal S}} \sum_{k=1}^K\mathbb{I}[\widecheck{y}_{n}^{(m)}=k]  \log [\A^{\bm \phi}_m(\bm x_n) \bm f_{\bm \theta}(\x_n)]_k +\beta \sum_{m=1}^M \sum_{n=1}^N {\rm trace}(\bm A_m^{\bm \phi} (\bm x_n)).
 \end{align}
However, establishing identifiability of the two functions $\bm A_m(\cdot)$ and $\bm f^\star(\cdot)$ from their product using E2E learning is fundamentally challenging.
} 
{{To establish identifiability of $\bm f^\star$ 
in the presence of instance-dependent noise, }\cite{nguyen2024noisy} considered a different {noise model}. There, the instance-dependent confusion is modeled as follows
\begin{align*}
\bm A_m(\bm x_n) = \bm A_m + \bm E_m(\bm x_n), \quad \forall m,n,
\end{align*}
with $\bm A_m$ being the annotator confusion component that is constant for all $n$, and the instance-independent perturbation $\bm E_m(\bm x_n) \neq \bm 0$ {only occurs occasionally---creating some outliers.} Under this {model,  \cite{nguyen2024noisy} showed that using $M\geq 2$ annotators can provably identify $\bm f^\star$ using an outlier-robust learning loss---interestingly reflecting that ``wisdom of the crowd'' is still useful in such challenging scenarios. 
}}

{To conclude our discussion on E2E crowdsourcing, we present a comparison of contemporary methods in Table \ref{tab:instance-dep} across three benchmark datasets. Among these, \texttt{COINNet} \cite{nguyen2024noisy} employs an instance-dependent label noise model, whereas \texttt{CrowdLayer}, \texttt{TraceReg}, \texttt{MaxMIG}, and \texttt{GeoCrowdNet} assume an instance-independent noise model. All approaches achieve competitive performance, with \texttt{COINNet} demonstrating superior results, likely due to its {more realistic noise model} and identifiability guarantees. 

}

}

{
\begin{table}[t]
    \centering
    \caption{Average classification accuracy on CIFAR-10N \cite{wei2022learning}, LabelMe \cite{russell2007labelme}, and ImageNet-15N \cite{nguyen2024noisy} datasets, labeled by human annotators. Top two results are bolded in black. Table is adopted from \cite{nguyen2024noisy}.}
    \label{table:cifar10n_labelme}
      \resizebox{0.65\linewidth}{!}{  \begin{tabular}{|c|c|c|c|}
    \hline
         \textbf{Method} & CIFAR-10N & LabelMe & ImageNet-15N \\ \hline \hline
\texttt{CrowdLayer} \cite{rodrigues2018deep}	&  87.38  $\pm $ 0.43  &  82.80 $\pm $ 0.90  &61.36	$\pm $ 2.73 \\
\texttt{TraceReg} \cite{tanno2019learning} &  86.57  $\pm $ 0.24  & 82.83 $\pm $ 0.23  &68.43 $\pm$ 0.12 \\
\texttt{MaxMIG} \cite{peng2019maxmig} 	& {\bf 90.11  $\pm $ 0.09} & 83.73 $\pm $ 0.84 & {\bf 81.13	$\pm $ 1.42} \\
\texttt{GeoCrowdNet(F)} \cite{ibrahim2023deep}	& 87.19 $\pm $ 0.37  & {\bf 87.21 $\pm $ 0.39} & 80.45	$\pm $ 1.77 \\
\texttt{GeoCrowdNet(W)} \cite{ibrahim2023deep}	& 86.43 $\pm $ 0.44  & 82.83 $\pm $ 0.75 &68.79	$\pm $ 0.27 \\ 
\texttt{COINNet} \cite{nguyen2024noisy} &\textbf{92.09 $\pm $ 0.47}	&\textbf{87.60 $\pm $ 0.54}& \textbf{93.71	$\pm $ 3.26} \\ \hline
    \end{tabular}}
    \label{tab:instance-dep}
\end{table}
}

\subsection{{Alternative} types of annotations}
\label{ssec:alternative_annotations}
While the discussions in Sec.~\ref{sec:label_integration} and \ref{sec:e2e} focus on the classification setting where the annotations are categorical, similar modeling and algorithm design ideas can be applied for other types as well. 

\noindent
{\bf Regression.} 
Consider the case where the ground-truth label and the corresponding annotations take continuous real values, {e.g., bounding box annotations in object recognition tasks, and  timestamp annotations in video activity recognition tasks. That is, we consider $y_n\in\mathbb{R}$ and $\widecheck{y}_n^{(m)}\in\mathbb{R}$ in such cases.} The most naive aggregation scheme is averaging, {i.e.,} \begin{equation*}
    \widehat{y}_n = \frac{1}{|\mathcal{M}_n|}\sum_{m\in\mathcal{M}_n}\widecheck{y}_n^{(m)}
\end{equation*}
with $\mathcal{M}_n\subseteq [M]$ denoting set of annotators who have provided a response for the $n$-th data item.  
A more robust alternative, that is often used in federated learning, is the median integration rule.

{Both average and median can be regarded as continuous extensions of majority voting.
The DS model can also be extended to continuous annotation cases.
}
For example, one can assign continuous distributions to the ground-truth labels and annotator responses, i.e.,   $y_n \sim \mathcal{N}(\mu_n,\sigma_n^2)$ where $\mu_n,\sigma_n^2$ are the mean and variance of the labels, respectively, while the conditional distributions of annotator responses given the ground-truth label $y_n = \alpha$ are  $\widecheck{y}_n^{(m)}|y_n=\alpha \sim \mathcal{N}(\alpha,\sigma_m^2)$, with $\sigma_m^2$ being annotator specific variance 
\cite{regression_ok19}. 
{A regression analogue of the DS model using the covariance matrix of the annotator responses are presented in \cite{tenzer2022crowdsourcing}, under the assumption that the responses have uncorrelated deviations given the ground truth. 
}
In the E2E learning-based setting, linear regression, Gaussian Process or deep learning models can be readily employed, with $\bm f_{\bm\theta}(\x_n)$ being the continuous distribution $ \prob(y_n | \bm x_n)$ \cite{rodrigues2018deep}. %

\noindent
{\bf Ranking.}
{In some cases, annotators are asked to rank data items. Two types of ranking are often considered, namely, pairwise preference and {ordinal ranking}.  

In the pairwise preference case, the {ordinal ranks} of all the data items are sought by comparing pairs. Given a pair $\bm{x}_n$ and $\bm{x}_{n'}$, the preference of $\x_n$ over $\x_{n'}$ is denoted as $\bm{x}_n \succ \bm{x}_{n'}$. One of the popular ways to model pairwise preferences is the {so-called} Bradley-Terry (BT) model, in which the probability of $\bm{x}_n \succ \bm{x}_{n'}$ is modeled as follows:
\begin{equation}\label{eq:preference_model}
    {\sf Pr}(\bm{x}_n \succ \bm{x}_{n'}) = \frac{e^{s_n}}{e^{s_n} + e^{s_{n'}}},
\end{equation}
where $s_n$ for $n\in [N] $ is a ranking score for $\x_n$. 
In the crowdsourcing setting, 
each annotator indicates their preference over $\bm{x}_n$ and $\bm{x}_{n'}$. That is, if annotator $m$ prefers $\bm x_n$ over $\bm x_{n'}$, it is denoted as $\bm{x}_n \succ_m \bm{x}_{n'}$. Given a ground-truth preference relation $\bm{x}_n \succ \bm{x}_{n'}$, annotator $m$'s correctness and confusion probabilities can be expressed using $w_m = \prob(\bm{x}_n \succ_m \bm{x}_{n'} | \bm{x}_n \succ \bm{x}_{n'})$ and $1-w_m$, respectively.  
Under this model, \cite{pairwise_ranking_crowds} used an MLE to jointly learn $w_m$ and ranking $s_n$. 
In E2E crowdsourcing, $s_n$ is expressed as $s_n=f_{\bm \theta}(\bm x_n)$, where $f_{\bm \theta}:\mathbb{R}^D\rightarrow \mathbb{R}$ learns a real-valued score function from the data features.
In the ordinal ranking case, \cite{raykar2010learning} converted the annotator-provided preference order of the data items into binary labels.
This also allows the associated MLE formulation to be combined with E2E learning method{---also see the inserted box ``Crowdsourcing for LLM alignment" that highlights the connections of these models and approaches to LLM alignment techniques.}

\noindent
{\bf Similarity annotations.} 
Crowdsourced E2E learning {can also be} formulated as a graph clustering problem. In this case, annotators are asked to indicate the similarity of two data items. This way, an (incomplete) $N\times N$ binary adjacency graph $\bm G$ is constructed, where $\bm G(i,j)=1$ means that $\x_i$ and $\x_j$ are regarded as ``similar'' or from the same class and $\bm G(i,j)=0$ means otherwise.
{Similarity-based annotation requires substantially lower expertise level from the annotators and thus is a promising paradigm for extra-large scale data annotation. }
The labels can be modeled as Bernoulli samples as follows:
\begin{align}\label{eq:bernoulli}
    \bm G(i,j)\sim {\rm Bernoulli}\left( \bm f^\star(\x_i)^\T \bm f^\star(\x_j) \right),
\end{align}
where $\bm f^\star$ such that $[\bm f^\star(\bm x_n)]_k = {\sf Pr}(y_n=k|\x_n)$ for $k=1,\ldots,K$, following the same realizability assumption before.
The term $\bm f^\star(\x_i)^\T \bm f^\star(\x_j)\in [0,1]$ naturally models the similarity of $\x_i$ and $\x_j$. 
Let a function $\bm f_{\bm \theta}$ be an approximation of the ground-truth $\bm f^\star$ as in the aforementioned E2E approaches. Then, the MLE under \eqref{eq:bernoulli} is a logistic regression objective (see \cite{nguyen2023deep} and some predecessors):
\begin{align}
    \minimize_{\bm f_{\bm \theta}\in \bm \varDelta_K} \sum_{(i,j)\in\bm \Omega} \left[  \mathbb{I}[{\bm G(i,j)}=1] \log \left( \bm f_{\bm \theta}(\x_i)^\T \bm f_{\bm \theta}(\x_j)  \right) + \mathbb{I}[{\bm G(i,j)}=0] \log \left( 1-\bm f_{\bm \theta}(\x_i)^\T \bm f_{\bm \theta}(\x_j)  \right)  \right],
\end{align}
where $\bm \Omega$ is the index set of annotated pairs.
The identifiability of $\bm f^\star$ can be established by treating the model in \eqref{eq:bernoulli} as a quantized nonnegative matrix factorization problem {in \cite{nguyen2023deep}}.
Similarity annotations can be easily acquired by the crowd (despite no annotator-specific models used in \eqref{eq:bernoulli}), and thus this type of approaches are often referred to as \textit{crowdclustering} in the literature.

\section{Emerging Topics in crowdsourcing}
As discussed in {earlier} sections, research in crowdsourcing has made impressive progress in the last several decades.
Next, we will explore some additional contemporary topics in this {field} that are {gaining traction} from the machine learning community.

 \subsection{Bias and fairness}
    \label{ssec:bias}
Machine learning and statistical learning algorithms are increasingly being applied in areas with significant human and societal impact, such as credit scoring, loan underwriting, job applications, and the penal system. While these algorithms often achieve overall good performance for the underlying task, they can also perpetuate unfairness and bias by discriminating against certain \textit{sensitive attributes}, such as race, age, or affiliation. In crowdsourcing, the presence of unfair workers, who provide more inaccurate labels to the data items belonging to a particular group, can significantly impact the overall system performance w.r.t. the sensitive attributes. {For instance, a recent empirical study \cite{lazier2023fairness} observed that the performance of the label integration approaches including majority voting, DS model-based EM, and the E2E approach Crowdlayer, all are impacted by the presence of unfair annotators. 
An approach to select fair workers for labeling tasks was advocated in \cite{goel2019crowdsourcing}, where an annotator assignment search algorithm was proposed. The assignment algorithm selects annotators such that the overall label integration accuracy is maximized while satisfying certain notions of fairness. Consider \textit{sensitive attribute-specific confusion matrix} per annotator, similar to the instance-dependent confusion matrix of \eqref{eq:insatnce_sig_model}, with entries { $ [\bm A_{m}(z_n=z)]_{k,k'} = {\sf Pr}(\widecheck{y}_n^{(m)}=k|y_n=k',z_n=z)$}
where $z_n \in \{0,1\}$ and $z_n$ is the random variable denoting the sensitive attribute (e.g., gender). Suppose that $s_m$ denotes the probability that any data item is assigned to $m$-th annotator, regardless of the sensitive attribute. Under this setting, \cite{goel2019crowdsourcing} sought an optimal annotator-assignment policy $\bm s= [s_1,\dots, s_M]$ by maximizing the expected labeling accuracy, i.e., $\sum_{z \in \{0,1\}} \prob(z_n=z) \sum_{k \in [K]} \prob(y_n=k) \sum_{m=1}^M s_m \bm [\bm A_{m}(z_n=z)]_{k,k}.$
Nevertheless, confusion matrices and the prior probabilities are assumed to be estimated \textit{a priori} using some limited ``gold" data items with known ground-truth labels and sensitive attributes. In addition, the maximization is performed under certain fairness constraints such as false positive rate parity on $\bm A_m(z_n=z)$'s and diversity constraints on $\bm s$ to balance the efforts across more annotators.}

\subsection{Adversarial attacks}
    \label{ssec:adversarial}
In crowdsourced settings, label noise sometimes stems from undesirable annotator behaviors rather than unintentional errors. Such undesirable behaviors may be caused by  annotators who provide random responses with no effort (spammers) and those who intentionally give incorrect responses (adversaries). 
In these cases, identifying and excluding spammers and adversaries is important for performance enhancement.

    {One approach to identify spammers is {through examining annotator confusion matrices}. Consider a spammer annotator $m$. Then, the entries of the corresponding confusion matrix are $\bm A_m(k',k) = \prob(\widecheck{y}_n^{(m)} = k' | y_n = k) = \prob(\widecheck{y}_n^{(m)} = k' | y_n = k'') = \prob(\widecheck{y}_n^{(m)} = k'), \forall k,k''$, i.e., the annotator response does not depend on the ground-truth label of the item. As such, all columns of $\bm A_m$ are the same and ${\rm rank}(\bm A_m)=1$.
    {Based on this observation, \cite{raykar_spammers} derived a ``spammer score'' that was then used in a modified EM algorithm to integrate labels and eliminate spammers. Using similar principles, a spectral approach based on the second-order moments $\left\{\mathbb{E}[\widecheck{\bm{y}}_n^{(m)}\circ\widecheck{\bm{y}}_n^{(m')}]\right\}$ can be used to identify spammers prior to label aggregation.}

    A particularly challenging setting involves colluding adversaries, i.e., annotators that cooperate to degrade the performance of a crowdsourcing system. Such a scenario can be particularly detrimental to label integration algorithms that rely on the conditional independence between annotators. A few recent works have attempted to tackle this issue. A robust rank-one matrix completion method was advocated in \cite{MMSR}.  Specifically, the rank-one plus identity structure, as given by  \eqref{eq:onecoin-ident} in { Sec.~\ref{ssec:methods_label_int}}, is exploited assuming that the reliabilities of the adversaries deviate from this structure. For the more general DS model, \cite{traganitis_adversaries} advocated a spectral method that leverages the particular structure of the annotator agreement matrix. 
    Note that the methods for dependent annotators in Sec. \ref{ssec:complex} attempt to unveil the dependencies between annotators. 
    Here, while there are dependencies between colluding annotators,  adversarial crowdsourcing algorithms treat adversaries as ``outliers'', and seek to filter them. 
    
    { When training deep learning systems, effective adversarial attack can be realized by adding designated noise to $\bm x_n$. Under such circumstances,} \cite{adv_learning_from_crowds} introduced an E2E method, based on EM, that can simultaneously learn annotator confusion matrices and a robust classifier.

    \subsection{Reinforcement learning with human feedback}
    \label{ssec:RLHF}
    A recent field gaining significant attention is \textit{reinforcement learning with human feedback} (RLHF), particularly for its application in fine-tuning LLMs~\cite{rlhf_survey_2024}.  In classical reinforcement learning (RL), an agent interacts with an environment over $T$ time steps. Per time step, the environment is at a state $s_t$ and the agent chooses an action $a_t$, and receives a reward $r_t.$ The next environment state $s_{t+1}$ is affected by the current state $s_t$ and the agents action $a_t,$ via a so-called transition probability $p(s_{t+1} | s_{t}, a_t).$ The goal of the agent is to learn a policy (i.e., a function mapping states to actions) that will maximize their rewards. RL has been extensively studied and a plethora of algorithms are available to train the agent. Nevertheless, a key component of RL is the reward function, whose design is a non-trivial task. RLHF circumvents the task of designing a reward function {\it a priori}; instead of using a pre-defined reward function, a reward model is learned via human feedback. In RLHF, the agent can issue queries to one or more humans and receive labels in response. Furthermore, queries can be specific actions and states, or trajectories $\tau$, which are sequences of states and actions $\tau = \{s_0,a_0,s_1,a_1,\ldots\}$. Typically, feedback is received asynchronously from the main RL agent-environment interaction. Human feedback can be in the form of classifying the queries as ``good'' or ``bad'', or as preferences between a pair of trajectories $(\tau_1,\tau_2)$. In the former case, the label integration approaches of the previous sections can be utilized to denoise the feedback from multiple human annotators. { The latter can be treated using preference-based learning as in \eqref{eq:preference_model}, and in RLHF settings $\x_n$ and $\x_n'$ represent two trajectories}.{ Indeed, the} Bradley-Terry model for RL {was} recently extended to the crowdsourcing case in \cite{crowd_prefrl}{---also see the inserted box ``Crowdsourcing for LLM alignment" for more recent advancements in this domain.}

        \begin{mdframed}
[backgroundcolor=gray!10,topline=false,
	rightline=false,
	leftline=false,
	bottomline=false]
    {
{\bf Crowdsourcing for LLM alignment}: Recent advancements in LLM research highlight the critical role of aligning models with human preferences. Methods such as RLHF \cite{zhao2023survey} and DPO \cite{rafael2023direct} have emerged as key paradigms for LLM alignment, often relying on crowdsourced {preference} annotations. DPO under noisy preference feedback from a single annotator {was} studied in \cite{chowdhury2024provably}, where human preferences are modeled using the Bradley-Terry (BT) model (see Eq. \ref{eq:preference_model}).
More recently, \cite{li2024aligning,chakraborty2024maxmin} extended RLHF to multi-annotator settings, aggregating diverse preferences through weighted majority voting, {showing promising performance enhancements.}
Overall, LLM alignment using crowd feedback remains an active research area with {exciting} open directions, including {preference noise modeling}, efficient learning algorithm {design}, and their performance characterization. Many of the approaches and models discussed in this manuscript share close connections to this problem and present promising avenues for future exploration.
}
\end{mdframed}

    \subsection{Connections to active learning, and transfer learning}
    \label{ssec:AL_TL}

    \subsubsection{Active learning}
    Most crowdsourcing jobs submitted to services such as AMT, operate within a predetermined budget. This budget can be easily translated into the total number of queries that can be asked from the annotators. Thus, efforts that reduce the overall amount of annotator queries, while maintaining classification performance are well motivated. { These goals can be formulated as an active learning problem}. 
    { Under a classification setup, active learning iteratively re-trains a learner by gradually selecting and adding samples to the training set in each iteration. Central components to any active learning system are the method that data is selected to augment the training set, and efficient retraining of the classifier as new data arrive. If these components are designed properly, the samples needed for learning the classifier can be drastically reduced relative to ordinary training processes. Data selection is typically facilitated by choosing data for which the current classifier is most ``uncertain'' of. Furthermore, { {(re-)training} $\bm f$} as new data arrive can be facilitated {via {online}} optimization methods such as stochastic gradient descent (SGD).}

    To minimize the number of queries to annotators, while maintaining high classification accuracy, active learning can be retooled and incorporated in the label integration procedure. 
    Notably, in addition to selecting which {data item} to acquire labels for, one must also select which annotators to query. Ideally, the annotators selected should be the ones with the highest probability of providing the correct label for a given {data item}. {At the same time, the set of selected annotators should be diverse, to prevent overloading the same annotators.}
    As discussed in Secs.~\ref{sec:label_integration} and \ref{sec:e2e}, popular label integration algorithms compute the posterior probability for each {data item} $\prob(y_n | \widecheck{\mathcal{Y}})$ at each iteration. These posteriors can then be used to quantify the ``uncertainty'' of the label integration for each {data item}, given currently available annotations. At the same time, the available estimates of annotator confusion matrices may not be reliable and introduce additional uncertainty into the model. Annotators can be selected using the current estimates of confusion matrices. However, one should be careful when selecting which annotator to query, as the ``best'' annotator based on current parameters may be suboptimal. 
    Thus, randomized exploration strategies may be beneficial, especially when only a few labels are available. 
    {A few recent methods utilize active learning in the pure label integration \cite{traganitis_active} and E2E cases~\cite{rodrigues2014gaussian}.}

    {In addition, it can be noted that the annotator selection problem can be cast as a Multi-armed Bandit (MAB) problem, with each annotator corresponding to an ``arm'' or ``action'' in the bandit setting \cite{MAB_w_online_estimation}.} In all these cases, {active learning}-based sampling of data and annotators outperforms randomly selected ones, indicating the significance of active methods in crowdsourcing.

    \subsubsection{Transfer learning}
    Another set of techniques designed to decrease the {amount of annotations are} { based on} {\it transfer learning} (TL). In TL, one is interested in training a model for a machine learning task of interest, termed the \emph{target task}. However, the target task does not have enough labeled data available for training. Instead, another related \emph{auxiliary task} has plenty of labeled data. TL seeks to \emph{transfer} knowledge from the auxiliary task, such that a high-performance model can be trained for the target task. In the crowdsourcing setup, annotators {can be regarded as ``vehicles''} to transfer knowledge across tasks. Indeed an interesting question that arises is: ``Which annotators would perform well on a { low-resource target} task?''.
    In \cite{cross_task_crowdsourcing}, a probabilistic model that captures annotator ability and task specific parameters was developed to estimate annotator reliability across multiple different tasks. When data features are available, \cite{crowdmtk} advocated for a logistic regression-based model that can transfer knowledge across multiple tasks. 

\section{Conclusions and future directions}

Crowdsourcing-based data annotation is pivotal in the AI era. Since crowdsourced annotators are often unreliable, 
{effectively integrating multiple noisy labels to produce accurate annotations stands as arguably the most important consideration for designing and implementing} a crowdsourcing system. In this feature article, we {reviewed} key milestones in crowdsourcing research, {including models, methods, theoretical advancements and emerging topics}.
{We also reviewed the intimate connections between crowdsourcing and signal processing theory and methods, such as NMF, tensor decomposition, distributed detection, and optimization techniques---showing how SP perspectives could offer principled design and enhanced performance for crowdsourcing systems.}

{Together with the AI boom and the wide adoption of} {large, complex models} that have high overfitting risks, the relevance of {crowdsourced data annotation} continues to rise. We highlight several worthwhile future research directions below:
\begin{itemize}
    \item Understanding of crowdsourcing under realistic and challenging scenarios, such as those involving instance-dependent label noise, imbalanced data distribution, or dynamic and evolving tasks, is still limited. Existing approaches either require stringent conditions or rely on heuristics. Principled solutions with performance guarantees—such as model identifiability, generalization, and sample complexity—are highly desirable in these cases.
    \item Crowdsourcing approaches are a key component of the training and fine tuning foundation models---{see {the box titled ``Crowdsourcing for LLM alignment"}}. {These directions have just begun to be explored, and thus many research questions, such as those regarding model building, method design, and performance characterization, remain wide open.}
    \item In-depth annotator behavior models, which consider aspects like bias and adversarial tendencies, have proven beneficial. However, these complex models often involve nontrivial and multi-faceted design considerations, such as fairness metric {effectiveness}, parameter {parsimony}, and algorithm {scalability}. {Despite some advancements in the past decade, unified design frameworks and standards have yet to be established.} 
    \item {Crowdsourcing offers solutions to a wide range of problems in science and engineering. However, there remains significant scope for the joint design of crowdsourcing algorithms and specific applications, such as medical diagnosis, environmental monitoring, and disaster response. This can be achieved by using cutting-edge models and algorithms tailored to the problem at hand, especially in light of the emergence of large-scale foundation models. Additionally, the collaboration between human and machine annotators, including LLM agents, can enhance data annotation. These opportunities present many unique challenges to be addressed in system design, data and computational resource management, and performance characterization.
    }

    \item {The scope of classical applications such as distributed detection, remote calibration, the CEO problem from information theory, and blind multichannel deconvolution can be broadened via cross-polination with crowdsourcing methods. For instance, blind deconvolution can be extended to nonlinear settings, using ideas from E2E crowdsourcing, whereas rate distortion ideas from information theory can be potentially applied to study the cost-performance tradeoff in crowdsourcing. Another exciting avenue of research include game theoretic approaches in adversarial crowdsourcing and robust data fusion techniques that learn from noisy multi-view data.  }

\end{itemize}

\section*{Acknowledgments}
{The work of X. Fu was supported in part by the National Science Foundation (NSF) under Project NSF IIS-2007836.} {The work of P. Traganitis was supported by NSF grant 2312546; {and the work of G. B. Giannakis was supported by  NSF grants 2126052, 2220292, 2312547, 2212318 and
2332547. Shahana Ibrahim and Panagiotis A. Traganitis contributed
equally to this work.}  }
\bibliographystyle{IEEEtran}
\bibliography{refs}

\end{document}

%% file: def.tex
\newcommand{\W}{\boldsymbol{W}}

\renewcommand{\H}{\boldsymbol{H}}

\newcommand{\A}{\boldsymbol{A}}

\newcommand{\x}{\boldsymbol{x}}
\newcommand{\f}{\boldsymbol{f}}

\newcommand{\w}{\boldsymbol{w}}

\newcommand{\T}{{\!\top\!}}

\newcommand{\tX}{\underline{\bm X}}
\newcommand{\tT}{\underline{\bm T}}

\def\diag{\mathrm{diag}}

\DeclareMathOperator*{\minimize}{\textrm{minimize}}
\DeclareMathOperator{\rank}{rank}

\DeclareMathOperator{\prob}{{\sf Pr}}
\DeclareMathOperator{\Expect}{\mathbb{E}}

\definecolor{orange}{RGB}{255,107,0}